\documentclass[lettersize,journal]{IEEEtran}
\usepackage{amsmath,amsfonts}
\usepackage{algorithm}
\usepackage{algpseudocode}
\usepackage{array}
\usepackage[caption=false,font=normalsize,labelfont=sf,textfont=sf]{subfig}
\usepackage{textcomp}
\usepackage{stfloats}
\usepackage{url}
\usepackage{verbatim}
\usepackage{graphicx}
\usepackage{cite}
\usepackage{booktabs} 
\usepackage{multirow}

\usepackage{colortbl}
\definecolor{mygray}{gray}{0.85}

\newtheorem{definition}{Definition}

\newcommand{\Ind}[1]{\hspace{#1ex}\hspace{#1ex}\hspace{#1ex}}

\newcommand{\swan}{\textsf{Swan}}
\newcommand{\verifix}{\textsf{VeriFix}}

\begin{document}

\title{VeriFix: Verifying Your Fix Towards An Atomicity Violation}

\author{Zhuang Li,Qiuping Yi,Jeff Huang,~\IEEEmembership{Member,~IEEE}

\thanks{This paper was produced by the IEEE Publication Technology Group. They are in Piscataway, NJ.}
\thanks{Manuscript received April 19, 2021; revised August 16, 2021.}}

\markboth{Journal of \LaTeX\ Class Files,~Vol.~14, No.~8, August~2021}%
{Shell \MakeLowercase{\textit{et al.}}: A Sample Article Using IEEEtran.cls for IEEE Journals}

\IEEEpubid{0000--0000/00\$00.00~\copyright~2021 IEEE}

\maketitle

\begin{abstract}

Atomicity violation is one of the most 
serious types of bugs in concurrent programs.
Synchronizations are commonly used to enforce atomicity.
However, it is very challenging to place synchronizations correctly and sufficiently
 due to complex thread interactions and large input space.
This paper presents \textsf{VeriFix}, a new approach for verifying
atomicity violation fixes.
Given a buggy trace that exposes an atomicity violation and
a corresponding fix, 
\textsf{VeriFix} effectively
verifies if the fix introduces sufficient synchronizations to repair the atomicity
violation without introducing new deadlocks.
The key idea is that \textsf{VeriFix} transforms the fix
verification problem into a property verification problem, in which both the
observed atomicity violation and potential deadlocks are encoded as a safety
property, and both the inputs and schedules are encoded as symbolic constraints.
By reasoning the conjoined constraints with an SMT solver, \textsf{VeriFix} 
systematically explores all reachable paths 
and verifies if there exists a concrete
\textit{schedule+input} combination to manifest the intended atomicity or
any new deadlocks.
We have implemented and evaluated \verifix\ 
on a collection of real-world C/C++ programs.
The result shows that
\textsf{VeriFix} significantly outperforms the state-of-the-art.

\end{abstract}

\begin{IEEEkeywords}
Atomicity violation, bug fix, dynamic analysis, verification.
\end{IEEEkeywords}
\section{Introduction}

\begin{figure}[t!]
\centering\includegraphics[scale=0.65, trim=275bp 300bp 280bp 150bp, clip]{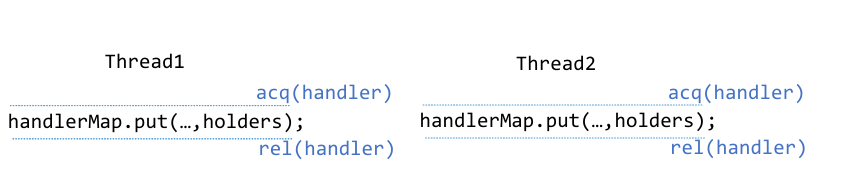}
\caption{SLING Bug-2812. The insufficient synchronization using a object field (handler), which is always different for different objects, to synchronize the codes. It will make the global object (handlerMap) broken. The incorrect fix for this issue wasn't identified until two years later.}
\label{fig:example}
\end{figure}

The difficulty of concurrent programming has led to a large number of techniques and tools to detect, repair, and verify concurrency errors such as race conditions and deadlocks~\cite{2017Speeding,Abdulla:SourceSet,Smaragdakis:2012,Jin:2011,
Liu:2012, Liu:2014, Liu:understand,Cai:17,Lin18,Costea21}.
While detecting concurrency errors is challenging, verifying their fixes is harder. Multiple studies~\cite{Liu:understand, Lu:2008} report that it can take months or even years to correctly fix a
concurrency bug, and nearly 70\% of the fixes are buggy in their first
try~\cite{Crispin:2000, Rescorla:2003}(Figure~\ref{fig:example} shows an incorrect fix, which was only discovered two years later).
The main problem is that the patches
generated by developers or automatic tools are typically only tested with the inputs exposing the original bug, but not been fully verified
with respect to different thread schedules and in particular alternative inputs.
More seriously, such unverified fixes could not only fail to fix a concurrency
bug, but also introduce new subtle bugs such as deadlocks.

Atomicity violations are among the most common concurrency bugs
~\cite{Liu:understand, Lu:2008},
which occur when the intended atomicity of a code region in one thread is
violated by interleaving accesses from other thread(s).
To fix an atomicity violation, the challenges are mainly two-fold:

First, fixing an atomicity violation requires global reasoning of the
whole program paths and schedules, which is essentially a difficult problem.
Typically, developers use synchronizations, such as locks, to ensure atomicity. However, due to the complex thread interactions and large input space,
it is often challenging to place locks sufficiently and correctly, resulting in
poor performance or deadlocks.

Second, most bug reports for atomicity violations only expose the currently
observed abnormal behavior, but not other parts of the target program that are unseen from the current execution.
A change that fixes the observed symptom does not necessarily fix the atomicity
violation bug, which may also be triggered by executing another function of the program.
Most existing automated fixing techniques for atomicity violation bugs~\cite{Jin:2011,
Liu:2012, Liu:2014, Liu:understand} heavily rely
on complete information from the bug report, which is usually unavailable.
Moreover, they do not systematically verify\footnote{Systematic verification involves exhaustively exploring all possible execution paths to ensure that the fix is effective not only for the current execution but also across the entire scheduling and input space of the program.} if the generated fix can correctly
enforce the intended atomicity along alternative paths or schedules given
different program inputs.

As an example, \swan~\cite{ShiHCX16} verifies if a
patch written in the form of Java synchronizations to an atomicity violation is sufficient for multithreaded Java programs.
It works by verifying a synchronization fix against an observed buggy trace and a set of thread schedules specific to the observed input.
However, \swan\ suffers from a strong limitation: it is sensitive to the observed buggy execution trace. That is,
it can only verify the fix under a specific input and a small space of alternative
schedules manifested in the observed trace. If the program runs under a
different input or a different schedule that is not captured by \swan,
the same atomicity violation may still occur. Moreover, a synchronization fix may introduce new deadlocks, which \swan\ cannot verify. 

\IEEEpubidadjcol

\begin{figure}[t!]
\centering\includegraphics[scale=0.35]{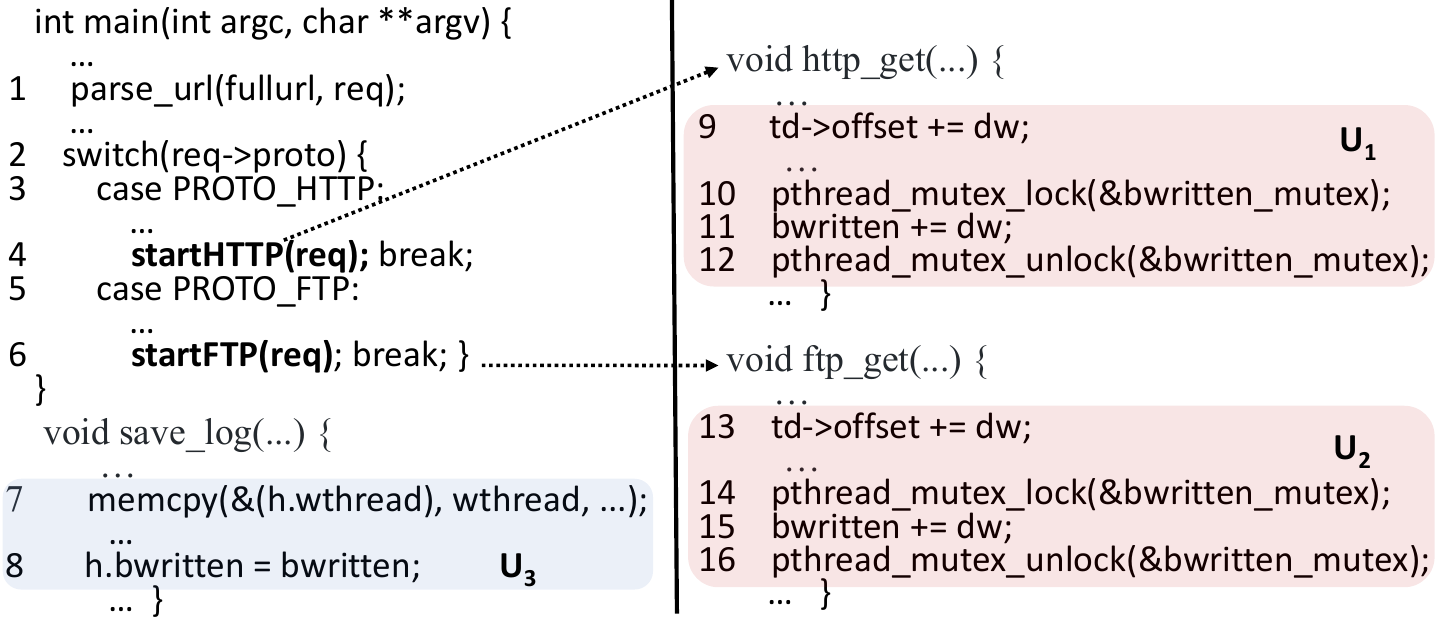}
\caption{An atomicity violation in \textit{Aget}, which
occurs when $U_3$ is interleaved by 
$U_1$ or $U_2$.}
\label{fig:aget}
\end{figure}

Consider an example in Figure~\ref{fig:aget}, which illustrates a real-world
atomicity violation in \textit{Aget}, a multi-threaded HTTP download
accelerator. When the user starts a downloading process with the HTTP protocol
the function \texttt{startHTTP} at line 4 will be invoked, which will execute
the work unit $U_1$
, and when the user stops the downloading
process (e.g., by typing \textit{ctrl-c} in the console), the work unit $U_3$
will be executed (to save the log).
The \texttt{memcpy}
function at line 7 saves the bytes downloaded by each thread, and the
assignment at line 8 saves the total number of bytes downloaded by all threads.
In the normal situation, the total number of bytes downloaded by each thread
should be the same as \texttt{bwritten}. However, the results will mismatch if
$U_3$ is interleaved by $U_1$ following a sub-trace {\tt 7-9-10-11-12-8}, which
manifests an atomicity violation.

To fix this atomicity violation, a straightforward way is to add a
\textit{lock-unlock} pair around $U_1$ and $U_3$ with a new lock.
However, this fix is in fact not sufficient because the atomicity violation
in $U_3$ can still be exposed to a different input.
Specifically, if the user starts a downloading process with the \texttt{FTP} protocol, 
the function \texttt{startFTP} will be invoked, which
will invoke \texttt{ftp\_get} and execute $U_2$. 
Because $U_2$ is not protected by the same lock, the atomicity violation in
$U_3$ remains.
Unfortunately, this fact cannot be identified based on the observed buggy sub-trace
{\tt 7-9-10-11-12-8}. 
Therefore, \swan\ 
will fail to validate that it is an incorrect fix.


In this paper, we present a new technique, \textsf{VeriFix}, that can
systematically verify atomicity violation fixes across different program
inputs and schedules. Unlike prior techniques such as \swan, we encode atomicity violations as safety properties and encode
{\em both} program inputs and schedules as symbolic constraints. 
By conjoining the constraints and reasoning them with an SMT solver, 
we are able to verify if the fix correctly enforces the intended atomicity 
in the {\em whole} scheduling and input space. 
Moreover, we can encode deadlocks as additional constraints 
in a similar way, 
such that we can also verify if the fix introduces new deadlocks or not.

A cornerstone of \verifix\ is systematic 
path exploration.
For each  path reachable from the whole schedule and input space, it
generates a concrete \textit{schedule+input} combination and uses a dynamic
scheduler to drive the program to execute the path. The trace along with each
path will produce new constraints, which are used not
only to verify the atomicity violation and deadlock properties, but also to
generate new combinations to explore new paths.

Another key characteristic of \verifix\ is that the verification can be 
performed in parallel, because each path exploration only relies on the generated
\textit{schedule+input} combination, and is independent of the other new paths
and explored paths. 
This feature allows \verifix\ to significantly improve 
efficiency through parallelization.




We implemented \verifix\ based on KLEE and Z3, where KLEE is used for executing the tested program and collecting constraints, while Z3 is employed to solve these constraints,
and evaluated it 
on a collection of real-world multithreaded C/C++ programs.
The results are highly promising: \verifix\ successfully detects all insufficient fixes which
cannot enforce the expected atomicity, and it identifies more atomicity violations and deadlocks than \swan.
We expect \verifix\ to be useful in several scenarios: 
First, during in-house development, when an
atomicity violation is encountered, developers would fix the program with
synchronizations. Then \verifix\ starts with the buggy trace and verifies
whether developers provide a sufficient fix without introducing 
deadlocks. Second, when a fix is generated by an
automatic fixing tool, \verifix\ can be used to systematically
verify whether the fix is both correct and sufficient. Third, \verifix\ can
be used to verify if the intended atomicity on a unit
of work is correctly enforced, even though currently no buggy trace has been
identified. For this scenario, developers can use the existing
synchronizations around the unit as a fix, and start \verifix\
with a random input and schedule, instead of a required buggy trace.

The key contributions of this paper are: 
 \begin{itemize}
 \item We present a new approach, \verifix, to verify if a
 synchronization fix to an atomicity violation is sufficient or not and 
 meanwhile ensure that it does not introduce new deadlocks.
 \item 
 We develop a systematic and
 parallel path exploration algorithm for concurrent programs that
 systematically explores all paths reachable from the whole schedule and input
 space through a constraint-based approach.
 \item We evaluate \verifix\ on a set of real-word C/C++ programs showing that 
 it significantly outperforms the state-of-the-art.
\end{itemize}


\section{Overview}
\label{sec:overview}

This section first illustrates the problem of verifying fixes for atomicity 
violations with our motivating example. 
Then, it presents an overview of \verifix\ and compares it with a state-of-the-art technique, i.e., \swan, in detail.

\subsection{Motivating Example}

First, the intended atomicity is not correctly enforced by 
\textit{fix1} because the two lines can still be interleaved by the remote write on $p$ 
at line 18 or 24 from thread $T_3$. 
For example, another trace $\tau^{'}$
$ = \{1-5, 5', 6, 16-18, 7\}$
can expose the same atomicity violation in 
the patched program. 
Nevertheless, 
it is difficult to find such a trace $\tau^{'}$, 
which requires a specific \textit{schedule+input} combination. 
To explore line 18, the branch at line 16 and 17 
should both take their true-branch, 
which requires the input $i = 2$, $j = 0$. 
Meanwhile, line 18 should be executed between lines 6 and 7. 
For an approach that explores different interleavings but with a fixed input, 
it will miss this atomicity violation. 
Second, \textit{fix1} introduces a deadlock among thread $T_1$, 
$T_2$ and $T_3$. 
Specifically, when the input $i$ equals to 2, it is possible to generate a lock-hold-wait cycle: 
$T_1$ holds lock $l_1$ and is waiting for lock 
$l_2$, $T_2$ holds lock 
$l_2$ and is waiting for lock $l_3$, and $T_3$ holds lock $l_3$ 
and is waiting lock $l_1$. 
Again, manifesting this deadlock is difficult, 
since it needs not only a complicated thread schedule but also a specific input.

Existing automatic fix generation techniques for 
atomicity violations cannot address the two problems above, 
because they do not take different inputs and schedules 
into consideration or do not consider deadlocks. 
In the rest of this section, we first illustrate how \swan~\cite{ShiHCX16}, 
a state-of-the-art atomicity violation fix verification technique, 
works on the example, and then show how \verifix\ systematically verifies the provided fix.


\begin{figure}[t]
\centering\includegraphics[scale=0.85]{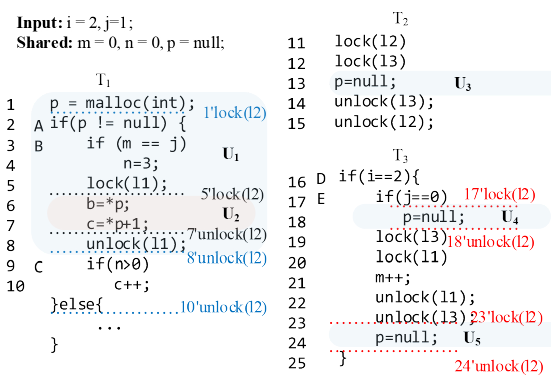}
\caption{
The motivating example with three fixes (\textit{fix1}—the black statements, \textit{fix2}—the blue statements, 
and \textit{fix3}—the blue and red statements). An atomicity violation between lines 6-7 can be 
exposed by a buggy trace $\tau_0 = \{16-17,19-25,1-3, 5-6, 11-15, 7\}$ with input $i = 2$, $j = 1$.
}
\label{fig:movitation}
\end{figure}

\subsection{\swan\ }
\label{swan}
  
Given a buggy trace containing a sequence of events that exposes an atomicity violation and a candidate fix in the form of a \textit{lock-unlock} pair, \swan\ leverages the reasoning capability of a constraint solver to check if the fix is sufficient in two steps. 
First, it detects a set of suspicious atomicity violations based on the observed buggy trace, 
by identifying all the other reads and writes 
in the trace that access the same memory location as that 
by the atomic region.
Second, it generates alternative schedules that can 
manifest those suspicious atomicity violations and
executes the patched program to follow 
such schedules at runtime attempting to expose those atomicity violations. 
If any atomicity violation is confirmed, \swan\ reports the fix is insufficient.

For our motivating example in Figure~\ref{fig:movitation} with 
the buggy trace $\tau_0$, \swan\ 
first generates trace $\tau_0^{'}$
$ = \{16-17,19-25,1-3, 5,$ 
$5^{'}$,
$6, 11-15, 7\}$ 
of the patched program with \textit{fix1} based on $\tau_0$. 
To ensure successful replay, when encountering the patched synchronizations, 
such as line $5^{'}$, \swan\ let the program skip them, 
but still record the synchronization events into the trace. 
Based on $\tau_0^{'}$, 
\swan\ identifies 4 suspicious atomicity violations: $(e_1, e_{13}, e_6), 
(e_1, e_{13}, e_7), (e_2, e_{13}, e_6), 
(e_2, e_{13}, e_7)$ 
where $e_i$ represents the event at line $i$. 
Then, it tries to expose all these six violations with 
enforcing event sequence
${e_1, e_2, e_{13}, e_6, e_7}$ 
in the patched program, and thus generates 
a trace $\tau_1 = \{16-17,19-25,1-3, 11-15, 5, 5^{'}, 6\}$. 
The patched program crashes along trace $\tau_1$ 
because the same null-pointer dereference is fired at 
line 6. Thus, \swan\ identifies that \textit{fix1} is insufficient to enforce the expected atomicity.

However, consider the change \textit{fix2} (the blue statements), 
which attempts to enforce atomicity between lines 2-8-10 with lock-unlock-unlock on $l_2$ 
(there are two unlock statements because of the branches), \swan\ will not be able to identify that \textit{fix2} 
is also insufficient, 
because no suspicious atomicity violation 
can be generated to expose the violation involving line~18 or line~24. 
Specifically, line~18 is not in the new $\tau_0^{'}$
$= \{16-17,19-25,1, 1^{'}, 2-3, 5-6, 11-15, 7\}$, 
generated for the patched program along the initial buggy trace $\tau_0$.
Furthermore, \swan\ only considers the relative order of $e_6$, $e_7$, and $e_{24}$ to be consistent 
with the bug trace $\tau_0$, so it will not identify ($e_6$, $e_{24}$, $e_7$) as a suspicious violation.
However, actually, the correct fix is \textit{fix3} 
(the blue and red statements).

\subsection{\verifix\ Overview}

\begin{figure}[t!]
\centering\includegraphics[scale=0.4]{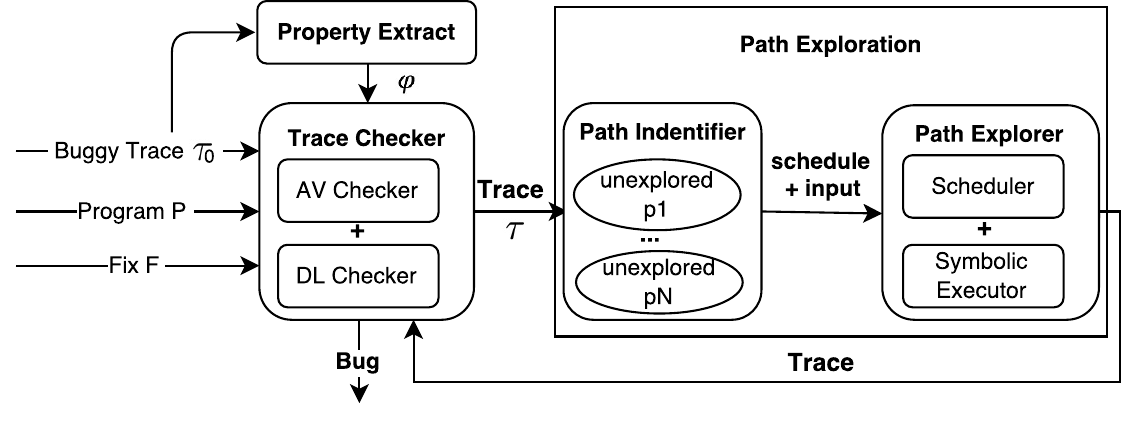}
\caption{Overview of \verifix.}
\label{fig:overall}
\end{figure}

Figure~\ref{fig:overall} depicts the overall flow of \verifix.
Given a fix and a corresponding buggy trace, \verifix\ first generates a trace $\tau$ 
by rerunning the patched program along with the same schedule and input in the 
given buggy trace. 
A random schedule is used when the actual execution does not follow 
the schedule of the buggy trace because of the synchronizations 
from the fix. If the atomicity violation does not manifest,  
then \verifix\ verifies the sufficiency
and correctness of the fix with an iterative process containing three main
components.
The first one, \textit{trace checker}, 
checks whether the fix provides sufficient synchronization to enforce the 
expected atomicity along the path manifested by trace $\tau$. 
The second one, \textit{path identifier}, identifies unexplored paths based on $\tau$ and 
generates a concrete \textit{schedule+input} component to trigger each unexplored path. 
The third one, \textit{path explorer}, invokes the scheduler and symbolic executor to 
explore the paths identified by the second component and then triggers the first component.

The three components iteratively explore new traces until an atomicity
violation bug or a deadlock is identified, or all reachable paths have been explored.
To fully verify a fix, \verifix\ will explore all reachable paths, 
though due to a large number of paths, it may not finish given a time budget.
In practice, we can run \verifix\ for a limited time.
If \verifix\ finishes, it proves that the fix has been fully verified to be
correct; if it runs timeout, although it does not prove the fix correct,
it means that the fix is correct on the explored paths, representing a large
schedule and input space.

\subsubsection{Verifying Atomicity Violations}


\verifix\ first translates the atomicity violation to a safety property.
For example, for the atomicity violation in Figure~\ref{fig:movitation}, it
constructs a property $\phi = (R_{p}^{6}=R_{p}^{7})$ based on the given buggy trace, where $R_{v}^{i}$
denotes the value read at line $i$ for the variable $v$.
The property $\phi$ requires that the read value at line 6 on $p$
equals to that at line 7, because the two reads are from the same atomicity
region without any intervening write on $p$, and are expected to return the same
value.
Now, the verification problem becomes to verify if there exists a path which
violates $\phi$.

\textsf{VeriFix} collects trace $\tau =$
$\{16-17, 19-25, 1-3, 5, 5^{'}, 6-7, 7^{'}, 8, 11-15\}$ 
based on the schedule and input which trigger buggy trace $\tau_0$ 
in the patched program with $fix_1$. 
Given $\phi$ and the trace $\tau$, 
\verifix\ then constructs a formula
$\Phi=\Phi_{rw}\wedge \Phi_{sync}\wedge\Phi_{pc}\wedge \neg\phi$.
The formula captures a thread causality model in which the variable $O_{i}$
denotes the order of an event $e_i$ (in $\tau$) in a possible schedule, which
also satisfies the data-validity constraints $\Phi_{rw}$, the thread
synchronization constraints $\Phi_{sync}$, and path constraints $\Phi_{pc}$.
We will explain these constraints in detail in Section~\ref{sec:approach}.
In particular, $\Phi_{pc}$ enforces the control-flow conditions along the path
due to both explicit branches and implicit path conditions such as $R_{p}^6\neq
null$. In our example, $\Phi_{pc}$ can be simplified as
$p = A \land \bar{B} \land P^6_p \land P^7_p \land \bar{C} \land D \land \bar{E}$, where $A$
and $\bar{A}$ respectively denote the branch condition of branch $A$ and its
negation (similarly for the other branches $B$, $C$, $D$ and $E$), and $P_v^i$ denotes the
constraint $R_{v}^i\neq null$.
Note also that $\Phi_{pc}$ also captures the program input $i$ and $j$ as
symbolic variables. Specifically, the branch conditions $\bar{B}$: $j \neq
R_{m}^{3}$ and $\bar{D}$: $i \neq 2$.


\verifix\ invokes an SMT solver on $\Phi$, which is satisfiable. 
Thus, there exists a \textit{schedule+input} combination that can 
violate the property $\phi$. Actually, the intended atomicity 
can still be interrupt by the remote write on $p$ at line 24 from thread $T_3$. 
Suppose the \textit{lock-unlock} pair on lock $l_2$ is added around $U_5$. 
Then \verifix\ will continue to verify $\phi$ along other reachable paths of 
the patched program by mutating the path conditions in $\Phi_{pc}$. 
Specifically, it tries to explore other unexplored path prefixes based on
$\Phi_{pc}$ (the details are described in Section~\ref{sec:pathprefix}), including $p_1: $ $A \wedge \bar{B}\wedge \bar{P_p^6}\wedge \bar{D}$, and
$p_2:$ $A\wedge B\wedge D$. 

For each new path, \verifix\ generates a concrete \textit{schedule+input}
combination from the solution returned from the solver.
For example, for $p_1$, the input $i=0$, $j=1$ is generated, and the
generated schedule requires line 13 to be executed after line 2 but before line
6. This combination is then enforced by \verifix\ to generate a new execution
of the program, which triggers a \textit{null}-pointer dereference at line 6,
exposing an atomicity violation between lines 2 and 6.

So far, \verifix\ identifies all the atomicity violations that \swan\ can. 
However, it does not stop but continues to explore other uncovered
paths such as $p_2$.
For checking $p_2$, it generates a new input $i=2$, $j=0$, which produces
trace $\tau_3 = \{1-5,5',6-7,7',8-23\}$. 
While verifying $\phi$ along $\tau_3$, \verifix\ generates a schedule with input $i=2$, $j=0$ 
to enforce line 18 to be executed between lines 6 and 7, 
Thus, the same atomicity violation is exposed again. 

\subsubsection{Verifying Deadlocks}
A correct fix should not only fix the atomicity violation but also ensure
that no new deadlocks are introduced. Different from \swan\ which may miss
deadlocks, \verifix\ also verifies that the provided fix, such as $fix_1$, 
does not introduce deadlocks along any reachable path.

Specifically, whenever a new trace is explored, \verifix\ constructs a
\textit{lock-event} graph (defined in Definition~\ref{def:legraph}), 
based on which it then identifies the
\textit{lock-hold-wait} cycles in the graph, which represent potential
deadlocks. For example, when trace $\tau$ is explored in the example, 
one such cycle involving all three threads is identified. 
The cycle describes the situation that thread $T_1$ holds $l_1$ and is waiting $l_2$, 
$T_2$ holds $l_2$ and is waiting $l_3$, and $T_3$ holds $l_3$ and is waiting $l_1$.

The potential deadlock needs to be confirmed by a concrete
\textit{schedule+input} combination. 
To generate such a combination, \verifix\ encodes the identified lock-hold-wait cycle 
into constraint $\phi = (O_{11} < O_{5'} \land O_{19} < O_{12} \land O_5 < O_{20})$, 
and checks the satisfiability of formula 
$\Phi =\Phi_{rw}\land \Phi_{sync} \land \Phi_{pc} \land \phi$. 
By solving $\Phi$, the solver generates a combination with input $i = 2$, $j = 1$ 
and schedule {$O_1 = 1, O_2 = 2, O_3 = 3, 
O_5 = 4, 
O_{11} = 5, O_{16} = 6, O_{17} = 7, O_{19} = 8$} to trigger the deadlock. 
The combination manifests the fact that the provided fix introduces a deadlock with the existing synchronizations. 
By enforcing the input and schedule, 
it will generate a trace with the deadlock, which can further help the developer 
understand the deadlock and guide the deadlock fixing process.


Overall, 
\verifix\ is able to identify the insufficient synchronizations and deadlock introduced 
by \textit{fix1} and the insufficient synchronizations introduced by \textit{fix2}.
\textit{Fix3} passes the verification after 7 paths are explored. 
Moreover, it reveals the following three facts in total. However, \swan\ can 
only reveal the first one.

\begin{enumerate}
\item Lines 2-7 should be in the same atomicity work unit $U_1$.
\item Line 18 and Line 24 cannot be concurrently executed with $U_1$.
\item The provided fix $fix_1$ not only fails to enforce the expected atomicity, but also
introduces a deadlock. 
\end{enumerate}


%


\section{The \verifix\ Approach}
\label{sec:approach}

%

Based on the observation that fixes for atomicity
violations in concurrent programs are prone to suffer
from insufficient synchronizations or introducing deadlocks,
 \verifix\ aims to address the problem by fully verifying the provided
 fixes. 
 Our work does not focus on detecting atomicity violations but assumes that a buggy trace exposing such a violation is already provided.
 More specifically, the verification problem is defined as follows:

\begin{definition}
\label{def:problem}
\textbf{(Fix Verification Problem ($\tau, U, M, L, F$)).}
Given a buggy trace $\tau$, which exposes an atomicity
violation on a code region $U$ while accessing memory locations $M$,
our approach verifies whether a fix $F$ in the form of adding/adjusting lock-unlock pair on a set of locks $L$ correctly enforces the intended atomicity on $U$ without introducing any new deadlocks.
\end{definition}


\verifix\ is a constraint-based approach that models program
executions, the expected atomicity, and potential deadlocks introduced by the
fix, all as symbolic constraints.
The fix verification problem is then formulated as a constraint solving
problem, in which a solver checks if there exists any execution that breaks the
expected atomicity or exposes a new deadlock.
Specifically,
let $\Phi_{\tau}$ be the execution constraints encoded from a trace $\tau$,
$\Phi_{av}$ and $\Phi_{dl}$ the expected atomicity constraints and deadlock constraints,
respectively, we check the satisfiability of two formulas $\Phi_1 =
\Phi_{\tau}\wedge \neg \Phi_{av}$ and $\Phi_2 = \Phi_{\tau}\wedge \Phi_{dl}$. 
If $\Phi_1$ is satisfiable then an atomicity violation is
identified, and if $\Phi_2$ is satisfiable then a deadlock is identified. 
Note that, $\Phi_{dl}$ encodes the deadlock constraint, and thus 
is not negated.  


\subsection{Constraint Encoding for One Trace}
\label{sec:encode}

\subsubsection{Encoding Execution Constraints}
\label{sec:enclodeTrace}

Inspired by existing work~\cite{Huang:MCR},
we encode a trace $\tau$ as a formula $\Phi_{\tau}$ over two types of variables:
\textit{O}-variables, denoting the order of critical events
in the trace, such as lock/unlock events, and \textit{R}-variables, denoting
the symbolic values of program inputs or reads.

$\Phi_{\tau}$ is the conjunction of three sub-formulas:
$\Phi_{\tau} = \Phi_{rw}\wedge \Phi_{pc}\wedge\Phi_{sync}$. $\Phi_{rw}$ denotes 
the data-validity constraints among
read / write events, and $\Phi_{pc}$ denotes the path conditions 
along $\tau$. $\Phi_{sync}$ denotes the synchronization constraints,
which is similar to that in~\cite{Huang:MCR}, except that it includes the
constraints introduced by the fix. We next focus on $\Phi_{rw}$ and $\Phi_{pc}$, which are different
from that in~\cite{Huang:MCR}.



Consider the read and write events on the same memory location
in the given trace.
A \texttt{Read} may {\em match} with a \texttt{Write} (i.e., \texttt{Read}
returns the value written by \texttt{Write}) by the same thread or from a
different thread, depending on the order relation between the \texttt{Read} and
the \texttt{Writes} and between the \texttt{Writes} themselves.
For a \texttt{Read} \texttt{r}, let \texttt{W} denote the set of
\texttt{Writes} on the same memory location as that of \texttt{r}, and $V_r$ the
value returned by \texttt{r}, then $\Phi_{rw}$ is defined as $\bigwedge_{r\in
\tau} \Phi_{rw}(r)$, where $\Phi_{rw}(r)$ is defined as follow:

\begin{equation}
\bigvee_{\forall w_{i}\in W} (V_r=w_i \wedge O_{w_{i}}<O_r \bigwedge_{\forall w_j\neq w_i} O_{w_{j}}<O_{w_{i}} \vee O_{w_{j}} > O_r)
\end{equation}

The above constraint states that if a \texttt{Read} is matched with
a \texttt{Write}, then the \texttt{Write}'s order $O_w$ must be smaller than
the \texttt{Read}'s order $O_r$ and at the same time there is no other \texttt{Write}
between them. The size of $\Phi_{rw}$, in the worst case, is cubic in the
size of the whole trace. 
(i.e., linear in the number of reads and quadratic in the number of writes).
Note that,
compared to the encoding in~\cite{Huang:MCR}, $\Phi_{rw}$ here is simpler in
that it does not contain constraints corresponding to the reachability of each
$w_i$, which is enforced implicitly by $\Phi_{pc}$.

$\Phi_{pc}$ is a conjunction of the path conditions for
all threads in the trace.
The path condition for each thread is conjunction of the explicit branch conditions
and implicit conditions on the dereferenced pointers (such as $R_n^1 > 0$
in our motivation example) along the path to ensure that all observed events 
are feasible. 

Each solution of $\Phi_{\tau}$ corresponds to
a concrete \textit{schedule+input} combination that can reach the same path as
$\tau$. Note that, different \textit{schedule+input}
combinations may cover the same path, so multiple solutions may exist for
$\Phi_{\tau}$.

\subsubsection{Encoding Atomicity Violations}
\label{sec:encodeAV}

\label{sec:encodeAV}

 \begin{table}[t!]
 \begin{center}
 \scalebox{.6}{
   \begin{tabular}{| l | l | l | c || l | l | l | c |}
     \hline
     \textbf{Type} & \textbf{Case}  & \textbf{Pattern} & \textbf{$\Phi_{av}$} & 
     \textbf{Type} & \textbf{Case}  & \textbf{Pattern} & \textbf{$\Phi_{av}$}\\\hline

     \multirow{4}{*}{} &
     \multirow{3}{*}{Case 1} & read$^p$ & & \multirow{4}{*}{} & \multirow{3}{*}{Case 5} & write$_{v1}^i$ & \\
	    & & \ \ \ \ \ \ \ \ \ \ $write^r$ &  $R^p == R^c$ & & & \ \ \ \ \ \ \ \ \ \ $write_{v1}^j$ &  $W_{v1}^i == R_{v1}^l\wedge$\\
	    & & read$^c$ & & & & \ \ \ \ \ \ \ \ \ \ $write_{v2}^k$ &  $R_{v2}^{i}==R_{v2}^{l}$ \\\cline{2-4}	    

  &   \multirow{3}{*}{Case 2} & write$^p$ &  & & & write$_{v2}^l$ & \\\cline{6-8}
	    &  & \ \ \ \ \ \ \ \ \ \ $read^r$ & $W^p \neq R^r$ &  &  \multirow{3}{*}{Case 6} & write$_{v1}^i$ & \\
S-V	& & write$^c$ & & & & \ \ \ \ \ \ \ \ \ \ $read_{v1}^j$ &  $R_{v1}^j \neq W_{v1}^i\wedge$ \\\cline{2-4}	    	    

  &	  \multirow{3}{*}{Case 3} & write$^p$ & & M-V  & & \ \ \ \ \ \ \ \ \ \ $read_{v2}^k$ &  $R_{v2}^k\neq R_{v2}^i$ \\
	    & & \ \ \ \ \ \ \ \ \ \ $write^r$ & $R^c == W^p$ &  & & write$_{v2}^l$ & \\\cline{6-8}
	    & & read$^c$ & & &   \multirow{3}{*}{Case 7} & read$_{v1}^i$ & \\\cline{2-4}

  &	 \multirow{3}{*}{Case 4} & read$^p$ & & & & \ \ \ \ \ \ \ \ \ \ $write_{v1}^j$ &  $R_{v1}^i == R_{v1}^l\wedge$  \\
	    & & \ \ \ \ \ \ \ \ \ \ $write^r$ & $R^p == R^c$ & & & \ \ \ \ \ \ \ \ \ \ $write_{v2}^k$ &  $R_{v2}^i == R_{v2}^l$ \\
	    & & write$^c$ & & & & read$_{v2}^l$ & \\\cline{2-4}

     \hline
   \end{tabular}}
 \end{center}
  \caption{Seven atomicity violation interleavings.
 S-V/M-V denotes single/multiple-variable atomicity violation.}
 \label{tab:fourcases}
 \end{table}

There are seven interleaving patterns for both single
and multiple variable atomicity violations~\cite{av-pattern}, as illustrated in
Table~\ref{tab:fourcases}.
For each pattern, we generate a corresponding constraint $\Phi_{av}$
denoting the intended atomicity property.
Specifically, let $W^i$ denote the value written by an event $write^i$, and
$R^i$ the \textit{old} value just before executing the event $write^i$. The
constraint $\Phi_{av}$ is constructed as follows:


\begin{itemize}
\item \textbf{Case 1}. Two \textit{local} read operations, $read^p$ and $read^c$, are interrupted by
a \textit{remote} write operation, $write^r$. $\Phi_{av}$ requires that
the two local reads must read the same value in all executions, 
as described by $R^p == R^c$. 

\item \textbf{Case 2}. Two \textit{local} write operations,
$write^p$ and $write^c$, are interrupted by
a \textit{remote} read operation, $read^r$.
$\Phi_{av}$ requires that 
the remote read $read^r$ should
not read value from the local write $write^p$, 
as described by $W^p \neq R^r$.

\item \textbf{Case 3}. A \textit{local} write operation
$write^p$ and a read operation $read^c$ are interrupted by
a \textit{remote} write operation $write^r$. 
$\Phi_{av}$ requires that 
$read^c$ always reads the value from
$write^p$, as described by
$R^c == W^p$.

\item \textbf{Case 4}. A \textit{local} read operation
$read^p$ and a write operation $write^c$ are interrupted by
a \textit{remote} write operation $write^c$.
$\Phi_{av}$ requires that
the \textit{old} value just before the event $write^c$ equals
to value $read^p$, as described by
$R^p == R^c$. 

\item \textbf{Case 5}. Two local writes on two variables
are interrupted by two remote writes on the same two variables.
To guarantee no writes are lost, $\Phi_{av}$ requires
$W_{v1}^i == R_{v1}^l\wedge R_{v2}^{i}==R_{v2}^{l}$.

\item \textbf{Case 6}. Two local writes on different variables
are interrupted by two remote reads. To ensure no temporary
result is seen to other threads, $\Phi_{av}$ requires
$R_{v1}^j \neq W_{v1}^i\wedge R_{v2}^k\neq R_{v2}^i$.

\item \textbf{Case 7}. Two local reads on different variables
are interrupted by two remote writes on the same two variables.
To guarantee the same memory state at two reads,
$\Phi_{av}$ requires $R_{v1}^i == R_{v1}^l\wedge R_{v2}^i == R_{v2}^l$.

\end{itemize}

\subsubsection{Encoding and Checking Deadlocks}
\label{sec:encodeDL}

Given an trace $\tau$, our basic idea is to construct a \textit{lock-event graph}
to detect potential deadlocks, and
encode them as a constraint (to be checked together with
the other constraints).

For convenience, we introduce three new notions. 
$l(e)$ represents the lock that the event $e$ requests,  
$h(\tau, t, i)$ represents the set of locks held by thread $t$ at position $i$
in $\tau$, and $event(\tau, t, i, lock)$ 
is the latest event which requires lock $lock$ before position $i$ in $\tau$ by thread $t$.
A lock can be requested more than once even in the same thread,
thus $event$ is vital for distinguishing different events which request the same lock.
$h(\tau, t, i)$ and $event(\tau, t, i, lock)$ can be abbreviated as $h(t)$ and $event(t, lock)$
respectively while traversing along $\tau$. 

\begin{definition}
\label{def:legraph}
(Lock-Event graph). Given an execution trace $\tau=\{e_1, e_2,...,e_n\}$. The
lock-event graph of $\tau$ is a directed labeled graph $G_{le}=(L, R, W, Event, T)$.
The node set $L$ represents locks, $Event$ represents the events in trace $\tau$,
and $T$ represents the thread ID set.
$W\subseteq T\times 2^{L}\times (Event\times Event)$ is the label set for edges, 
and $R\subseteq L\times W\times L$ represents the edge set. 
An edge $(l_1, (t, h(\tau,t, i), \langle e_x, e_i\rangle), l_2)$ between the lock $l_1$ and $l_2$ exists,
\textit{iff} 
$e_i$ requests $l_2$ (i.e., $l(e_i) = l_2$), 
$l_1$ is hold by thread $t$ before $e_i$ (i.e., $l_1 \in h(\tau, t, i)$), and the event $e_x$ is
the latest event which requests $l_1$ (i.e., $e_x=event(\tau, t, i, l_1)$). 
\end{definition}

\begin{algorithm}[t!]
  \caption{{\bf PotentialDLs: construct the lock-event graph for trace $\tau$ and return
  unsafe cycles.}}
  \label{alg:potentialDL}
{\footnotesize
\begin{algorithmic}[1]
  \State{\Ind{0} \textbf{Input:} trace $\tau$;}
  \State{\Ind{0} \textbf{Output:} potential deadlocks $deadlocks$;}
  \State{\Ind{0} $h(t)\leftarrow \emptyset$;}
  \State{\Ind{0} {\bf for} (each event $e_i$ in $\tau$)}
  \State{\Ind{1} {		\bf if} ($e_i$ requires lock $l$ in thread $t$)}
  \State{\Ind{2} {			\bf for} (each lock $l'\in h(t)$)}
  \State{\Ind{3} {			$G_{le} = G_{le}\cup \{(l', (t, h(t), \langle event(t, l'), e_i\rangle), l)$;}} 
  \State{\Ind{2} {			$h(t) = h(t)\cup \{l\}$, $event(t, l) = e_i$;}}
  \State{\Ind{1} {		\bf else if} ($e_i$ requires unlock $l$ in thread $t$)}
  \State{\Ind{2} {			$h(t) = h(t)\setminus \{l\}$;}}
  \State{\Ind{0} {\bf for} (each cycle $c$ of $G_{le}$ containing at least one lock in $L$)}
  \State{\Ind{1} {		Let $W=\{w_1, ..., w_n\}$ be the edge labels in $c$}}
  \State{\Ind{1} {		\bf if} ($\forall w_i\neq w_j\in W$, $(T(w_i)\neq T(w_j))\wedge (h(w_i)\cap h(w_j)=\emptyset)$)}
  \State{\Ind{2} {			$list\leftarrow \emptyset$;}}
  \State{\Ind{2} {			\bf for} (each edge $(l_i, (t, h, \langle e_x, e_y \rangle), l_j)$ in $c$)}
  \State{\Ind{3} {				$list.add(\langle e_x, e_y \rangle)$;}}
  \State{\Ind{2} {			$deadlocks.add(list)$;}}
  \State{\Ind{0} {	\bf return} $deadlocks$;}

\end{algorithmic}
}
\end{algorithm}

For an edge label $w$, we use $T(w)$ and $h(w)$ to represent
its thread ID set and the $h$ set. 
Algorithm~\ref{alg:potentialDL} describes the process of identifying the potential
deadlocks based on $\tau$. 
It first constructs $G_{le}$ between
line~3 $\sim$ 10. Specifically, it mainly updates $h(t)$ and $event(t, lock)$,
where $h(t)$ maintains the currently hold locks by thread $t$ and $event(t, lock)$ represents the last event which
requests $lock$ by thread $t$.
Line 11 $\sim$ 17 identifies the potential deadlocks in $G_{le}$,
by computing the \textit{unsafe} cycles (Definition~\ref{def:unsafecycle}).
Each potential deadlock is a list of event pairs, $\{\langle e_1, e_2\rangle,
...\langle e_{n-1}, e_n\rangle\}$. 

\begin{definition}
\label{def:unsafecycle}
(An Unsafe Cycle). A cycle $c$ with edge label set $W=\{w_1, w_2, ..., w_n\}$ in $G_{le}$ is unsafe \textit{iff} it satisfies
$\forall w_i\neq w_j\in W$, $(T(w_i)\neq T(w_j))\wedge (h(w_i)\cap h(w_j)=\emptyset)$. 
\end{definition}

\begin{algorithm}[t!]
  \caption{{\bf CheckDL: check deadlock based on a given trace.}}
  \label{alg:checkdl}
{\footnotesize
\begin{algorithmic}[1]
  \State{\Ind{0} \textbf{Input:} trace $\tau$;}
  \State{\Ind{0} \textbf{Output:} schedule+input combination $sch+inp$;}

  \State{\Ind{0} $deadlocks\leftarrow $ \bf PotentialDLs $(\tau)$;}
  \State{\Ind{0} {\bf for} (each $dl$ in $deadlocks$)}
  \State{\Ind{1} {		$\Phi_{dl} = True$;}}
  \State{\Ind{1} {		Normalize $dl$ to $\{\langle e_1, e_2\rangle, ...\langle e_{n-1}, e_n\rangle\}$;}}
  \State{\Ind{1} {		\bf for} (each $i\in\{1, n/2-1\}$ )}
  \State{\Ind{2} {			$\Phi_{dl} = \Phi_{dl}\wedge (O_{e_{2i}} > O_{e_{2i+1}})$;}}
  \State{\Ind{1} {		$\Phi_{dl} = \Phi_{dl}\wedge (O_{e_{n}} > O_{e_{1}})$;}}
  \State{\Ind{1} {		$sch+inp\leftarrow solve(\Phi_{\tau}\wedge \Phi_{dl})$}}
  \State{\Ind{1} {		\bf if} ($sch+inp\neq NULL$)}
  \State{\Ind{2} {			\bf return $sch+inp$};}
  \State{\Ind{0} {\bf return $NULL$;}}
\end{algorithmic}
}
\end{algorithm}

\begin{figure}[t!]
\centering\includegraphics[scale=0.4]{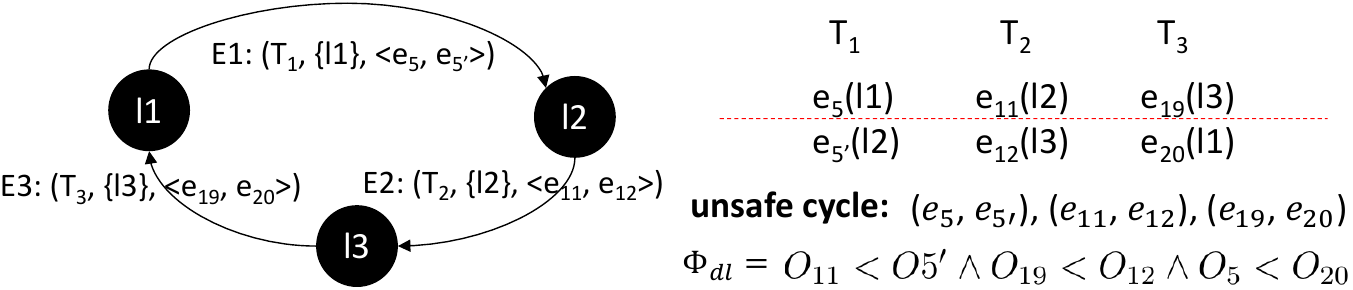} 
\caption{The lock-event graph for the trace $\tau_{0}$ in Figure~\ref{fig:movitation}.}
\label{fig:cycle}
\end{figure}

Algorithm~\ref{alg:checkdl} presents the process of checking deadlocks along an
observed trace $\tau$. It first identifies the potential deadlocks at line~3 by invoking \textit{PotentialDLs}
(Algorithm~\ref{alg:potentialDL}). Then, for each potential deadlock $dl$, it generates
the corresponding formula $\Phi_{dl}$.
Specifically, it normalizes $dl$ to $\{\langle e_1, e_2\rangle, ...\langle e_{n-1}, e_n\rangle\}$,
where $(\forall i\in\{1, n/2-1\}, l(e_{2i})==l(e_{2i+1}))\wedge (l(e_{n})==l(e_1))$. Then it constructs $\Phi_{dl}$ as
$\bigwedge_{i\in\{1, n/2-1\}}(O_{2i}>O_{2i+1})\wedge (O_{n} > O_{1})$, as described between
line~7$\sim$9. Note that, $\Phi_{dl}$ describes the constraints that deadlock happens, thus a SMT-solver
is invoked at line 10 to generate and return a \textit{schedule+input} combination to trigger the deadlock at line 12. 
$NULL$ is returned when no deadlock is confirmed.

For our motivating example, Figure~\ref{fig:cycle} shows the only \textit{unsafe}
cycle identified by Algorithm~\ref{alg:potentialDL},
encoded as $\Phi_{dl}$. 
By checking $\Phi_{\tau}\wedge\Phi_{dl}$,
we get a solution with input $i=2$, $j=1$ and 
a schedule which requires line 11 is executed after line 5 but before line 5’, 
and line 19 is executed after line 11 but before line 12. This combination can be used to trigger the deadlock.

\subsection{Fix Verification Based on Path Exploration}
\label{sec:wholeAlg}

\subsubsection{The Overall Algorithm}
\label{sec:pathexplore}




\begin{algorithm}[t!]
  \caption{{\bf Atomicity Violation Fix Verification}}
  \label{alg:basic}
{\footnotesize
\begin{algorithmic}[1]
  \State{\Ind{0} \textbf{Input:} Program $P$, Fix $F$, buggy trace $\tau$;}
  \State{\Ind{0} \textbf{Output:} \textit{schedule+input} combination $s+i$;}
  \State{\Ind{0} $P'\leftarrow$ the patched program on $P$ with $F$; }
  \State{\Ind{0} $(s_0+i_0) \leftarrow$ The \textit{schedule+input} combination
  which reproducing buggy trace $\tau$ in Program $P$;}
  \State{\Ind{0} $workList.insert(\langle s_0+i_0, True\rangle)$;}
  \State{\Ind{0} {\bf while} ($workList\neq \emptyset$)}
  \State{\Ind{1} {$\langle s+i, prefix\rangle\leftarrow workList.pop()$;}}
  \State{\Ind{1} {$\tau \leftarrow \bf GuidedSE$ $(s+i, prefix, P')$;}}
  \State{\Ind{1} {$s+i\leftarrow$ \bf checkAV $(\tau)$;}}
  \State{\Ind{1} {\bf if} ($s+i\neq NULL$)}
  \State{\Ind{2} {		\bf Report an atomicity violation bug; \bf return $s+i$;}}
  \State{\Ind{1} {$s+i\leftarrow$ \bf checkDL $(\tau)$;}}
  \State{\Ind{1} {\bf if} ($s+i\neq NULL$)}
  \State{\Ind{2} {		\bf Report an deadlock bug; \bf return $s+i$;}}
  \State{\Ind{1} {$list\leftarrow$\bf GenerateNewSI($\tau, prefix$);}}
  \State{\Ind{1} {$workList.insert(list)$;}}
  \State{\Ind{0} {\bf return $NULL;$}}
\end{algorithmic}
}
\end{algorithm}

Algorithm~\ref{alg:basic} shows our overall fix verification algorithm.
It takes a program $P$, a buggy trace $\tau$ of $P$, and a fix $F$ as
input and returns either $NULL$ meaning that the fix is verified without
finding any atomicity violation or deadlock, or a \textit{schedule+input}
combination ($s+i$), which can trigger a identified atomicity violation or deadlock.

Specifically, it uses $workList$ to maintain a list of pairs $\langle s+i,
prefix \rangle$, where $s+i$ is a \textit{schedule+input}
combination and $prefix$ a path prefix.
Each $\langle s+i, prefix\rangle$ can be used to drive an execution of the
patched program $P'$ (constructed at line 3) that follows the path prefix $prefix$ with
input $i$ and schedule $s$.
The $workList$ is initialized with $\langle s_0+i_0, True\rangle$, denoting the
schedule and input of the original buggy trace and an empty path prefix.
The prefix $True$ states that no path has been checked at the beginning. 
Note that, a random schedule is used when the actual execution does not follow 
the schedule $s_0$ because of the synchronizations from the fix.

For each element $\langle s+i, prefix\rangle$ in $workList$, 
$GuidedSE$ at line 8 generates a symbolic trace $\tau$
of the patched program $P'$ with the guide of $s+i$.
Specifically, it collects path conditions and records critical events, such as
lock/unlock events, thread start/joint events, and read/write events on shared
variables.
$\tau$ is then used to check atomicity violations and deadlocks at lines 9 and
12, respectively, based on the constraint encoding as described in
Section~\ref{sec:encodeAV} and 
Section~\ref{sec:encodeDL}. If an atomicity violation or a deadlock is found, the algorithm returns with a
corresponding bug triggering $s+i$ combination.
If not, the algorithm continues to explore new paths.
Based on $\tau$ and the current path prefix $prefix$,
\textit{GenerateNewSI} at line 15 (described next in
Section~\ref{sec:pathprefix}) is used to generate new $\langle s'+i',
prefix'\rangle$ items that can trigger new unexplored path prefixes.
The new items are added to the $workList$ and then the algorithm is repeated.
Algorithm~\ref{alg:basic} stops when a bug is found or 
$workList$ becomes empty, meaning that no unexplored path can be
identified.

\begin{algorithm}[t!]
  \caption{{\bf GenerateNewSI}($\tau$, $prefix$)}
  \label{alg:gen}
{\footnotesize
\begin{algorithmic}[1]
  \State{\Ind{0} $SI \leftarrow \emptyset$};
  \State{\Ind{0} Let $p$ be the path of $\tau$;}
  \State{\Ind{0} {\bf for each} {$pre'$ $\in$ $split(pre, p)$}}
  
  \State{\Ind{1} $\tau' \leftarrow$ extractSubTrace($\tau$,$pre'$);}
  \State{\Ind{1} $\Phi_{pc}\leftarrow \bigwedge _{c\in pre'} c$;}
  \State{\Ind{1} $\Phi_{sync} =$ SynchronizationConstraints($\tau'$);}
  \State{\Ind{1} $\Phi_{rw} =$ RWConstraints($\tau'$);}
  \State{\Ind{1} $\Phi_{pre'}=\Phi_{pc}\wedge \Phi_{sync}\wedge \Phi_{rw}$;}
  \State{\Ind{1} $si \leftarrow$ \bf{solve}($\Phi_{pre'}$); }
  \State{\Ind{1} {\bf if} ($si \neq null$) {\bf then}}
  \State{\Ind{2} $SI.add(si, pre')$;}
   \State{\Ind{0} {\bf Return $SI$.}}
\end{algorithmic}
}
\end{algorithm}
\subsubsection{Path-prefix Guided Exploration}
\label{sec:pathprefix}

  

The procedure \textit{GenerateNewSI} (shown in Algorithm~\ref{alg:gen}) is an
efficient path exploration algorithm that, guided by the observed
trace and a path prefix, identifies new paths to explore.

Consider a trace $\tau$ with two threads driven by a path prefix
$pre$. Suppose $pre$ is extended with suffixes $A$ and $B$ for each of the
two threads, respectively. That is, the newly explored path $p$ is $pre$-$A$-$B$.
Note that $A$ and $B$ are path conditions with respect to two branch choices.
Hence, there are three
new possible path prefixes identified by combining different branch choices:
$pre$-$\overline{A}$-$B$, $pre$-$A$-$\overline{B}$ and
$pre$-$\overline{A}$-$\overline{B}$, where $\overline{X}$ means
the negation of $X$, i.e., the path follows the opposite choice of $X$.


More generally, let $split(pre, p)$ refer to the set of newly identified path prefix
combinations based on a newly explored path $p$ and a path prefix
$pre$, and let $\textit{suffix}(T_i)$ denote the path extension for thread $T_i$ from
$pre$ to $p$, i.e., removing $pre(T_i)$ from $p(T_i)$.
Then $split(pre, p)$ contains new path prefixes formed by all combinations of
$\textit{suffix}(T_i)$ and its negations among all threads.
For each individual thread, $\textit{suffix}(T_i)$ may exhibit more than
one new branch.
For example, suppose $\textit{suffix}(T_i)$=$b1$-$b2$-$b3$, then three
path prefixes are needed to be explored: $\overline{b1}$, $b1$-$\overline{b2}$
and $b1$-$b2$-$\overline{b3}$. Note that other
combinations such as $\overline{b1}$-$\overline{b2}$-$\overline{b3}$ are not
valid because the execution of a branch may depend on the preceding branch
choices.
In total, $split(pre, p)$ contains
${\displaystyle\prod^{N}_{i=1}(\vert \textit{suffix}(T_i)\vert +1)-1}$ newly 
identified path prefixes, 
where $N$ is the number of threads and $\vert \textit{suffix}(T_i)\vert$ the number of branches
in $\textit{suffix}(T_i)$. 

For each newly identified path prefix $pre'\in split(pre, p)$, Algorithm~\ref{alg:gen} first
extracts sub-trace $\tau'$ guaranteed by $pre'$ in $\tau$, and then constructs
execution constraints along $\tau'$ following Section~\ref{sec:enclodeTrace}.
If the constraints are satisfiable, it generates a
corresponding \textit{schedule+input} combination that can drive the program to
follow $pre'$.

\subsubsection{Parallelizing Path Exploration}
Different from other state-space exploration techniques~\cite{Flanagan:POR,Coons:2013,Musuvathi:ICB2}
which are hard to parallelize, Algorithm~\ref{alg:basic} can be easily
parallelized. Specifically, the online path exploration can be parallelized because it only
depends on the provided \textit{schedule+input} combination.
Similarly, the offline path identification can also
be parallelized, because it only depends on the trace information collected
from the observed execution.

Algorithm~\ref{alg:parallelVeriFix} describes the parallelized \textsf{VeriFix} algorithm.  
It can be started by invoking \textsf{Parallel-VeriFix}$(s_0+i_0, True)$, where $s_0+i_0$ represents 
the \textit{schedule+input} combination of the provided buggy trace. 
For each $si$ and the corresponding path prefix,
\textsf{Parallel-VeriFix} first carries out a guided symbolic execution to get trace 
$\tau$ of the patched program at line 1 as in the sequential algorithm, and then checks atomicity violations and 
deadlocks along trace $\tau$ at line 2 and line 6 respectively. If no bug is identified, it 
then invokes \textit{Parallel-GenerateNewSI} to
generate new $si$ list SI and path-prefixes in parallel.
For each new $si$ and path prefix $p$, it immediately starts a parallel instance of
\textsf{Parallel-VeriFix} at line 12 to continue exploring. 
Furthermore, a number $N$ can be used to control its parallelism.
Thus, it parallelizes a new path exploration process only when less than $N$ such processes
are still running, otherwise, stores the generated items to $workList$ as the sequential algorithm. 

\begin{algorithm}[t!]
  \caption{{\bf Parallel-VeriFix($si$, $prefix$)}}
  \label{alg:parallelVeriFix}
{\footnotesize
\begin{algorithmic}[1]
  \State{\Ind{0} {$\tau \leftarrow \bf GuidedSE$ $(si, prefix)$;}}
  \State{\Ind{0} $s+i\leftarrow$ \textbf{checkAV}$(\tau)$;}
  \State{\Ind{0} {\bf if} ($s+i\neq NULL$)}
  \State{\Ind{1} {		\bf Report an atomicity violation bug;}}
  \State{\Ind{1} {		\bf return $s+i$;}}
  \State{\Ind{0} $s+i\leftarrow$ \textbf{checkDL}$(\tau)$;}
  \State{\Ind{0} {\bf if} ($s+i\neq NULL$)}
  \State{\Ind{1} {		\bf Report an atomicity violation bug;}}
  \State{\Ind{1} {		\bf return $s+i$;}}
  \State{\Ind{0} $SI \leftarrow$ {\bf Parallel-GenerateNewSI($\tau, prefix$): }}
  \State{\Ind{0} {\bf par-forall} {($si, p$) $\in$ $SI$}}
  \State{\Ind{1} {{\bf Parallel-VeriFix}($si, p$);}}
\end{algorithmic}
}
\end{algorithm}

%
%
%

\section{Evaluation}
\label{sec:eval}

We evaluated our approach with four sets of experiments. 
First, we compared \verifix\ and \swan\ with fixes 
of different qualities on several small programs. 
Second, we compared them on a series of incorrect fixes in real-world programs. 
Third, we compared them on a series of fixes generated by 
three representative concurrent bug repair tools, 
including $\alpha$Fixer~\cite{Cai:17}, PFix~\cite{Lin18}, Hippodrome~\cite{Costea21}. 
At last, we evaluated \verifix\ on a series of correct fixes in real-world programs by 
comparing with Deagle~\cite{Deagle2022}, the state-of-the-art concurrent verification tool based on CBMC for verifying program security properties using constraint solving.

However, none of the above tools have deadlock detection capabilities. Therefore, when addressing deadlocks, we compare our approach with the state-of-the-art tool SPDOffline~\cite{deadlock-1}. Since SPDOffline relies on traces provided by other tools and can only predict whether a trace contains a reordering that might trigger a deadlock, our comparison with SPDOffline is limited to our deadlock detection module, specifically Algorithm~\ref{alg:potentialDL} and Algorithm~\ref{alg:checkdl} of \verifix. We manually provide SPDOffline with a trace identified by \verifix\ that contains potential deadlocks and then compare whether SPDOffline and our deadlock detection module can detect deadlocks in that trace.

Correspondingly, we aim to answer these research questions:
 \begin{itemize}
        \item RQ1: How efficient and effective is \verifix\ in fix verification across both the schedule space and the input space?
	\item RQ2: How efficient and effective is \verifix\ in 
            exposing insufficient fixes of real-world programs? 
	\item RQ3: How efficient and effective is \verifix\ 
            in verifying the fixes generated by Automated program repair (APR) techniques for real-world programs? 
	\item RQ4: How scalable is \verifix\ in verifying sufficient fixes of real-world programs? 
 \end{itemize}

To present our experimental findings in a simple and clear manner, we did not show the results of VeriFix's parallel algorithm when answering RQ1 to RQ3. This is because our parallelization algorithm does not contribute to improving the efficiency of the evaluation for RQ1 to RQ3; its performance is similar to the non-parallelized approach. The parallelization primarily takes effect after the path identifier generates multiple path prefixes, allowing our parallel algorithm to launch multiple path explorers to execute these path prefixes simultaneously.

We ran all our experiments on a sixteen-core Ubuntu 20.01 machine
with 2.90GHz Intel(R) Xeon(R) Gold 6226R CPU, and set a timeout of 20 minutes for
each benchmark. In addition, we set the maximal
parallelism of \verifix\ to 5 (i.e., we allow five concurrent threads at most),
and set the loop unfold depth to 5.

\subsection{Implementation}

We implemented \verifix\ for C/C++ programs on top of CLAP~\cite{Huang:2013}, which is built upon KLEE and Z3, where Z3 serves as the constraint solver used by KLEE.
The tool first reproduces the buggy trace using CLAP 
and then constructs the properties and verifies them in the following iterative process.
\verifix\ contains three components: trace checker,
path identifier, and path explorer, as depicted in Figure~\ref{fig:overall}.
Whenever a new path is explored, the trace checker checks whether the 
provided fix provides sufficient synchronization on the identified 
atomicity violation without introducing deadlocks. 
The path identifier takes the just checked trace as input, constructs the
corresponding constraint model, and invokes Z3 to identify newly unexplored paths. 
The solution corresponds to a list of \textit{schedule+input} combinations 
that trigger the iterative verification process. 
The path explorer uses a special scheduler to enforce the thread execution to follow 
each \textit{schedule+input} combination.
It also contains a symbolic tracer, which performs symbolic execution along 
the controlled execution and tracks its path conditions and the trace information.

\verifix\ aims to explore all reachable paths, which is inherently 
challenging for programs with loops. To address this problem, 
we provide a configuration that allows the unfolding of each loop in 
the program a limited number of times. In our evaluation, we 
set the loop unfolding depth to 5. 
To compare with the state-of-the-art, we also implemented \swan\ 
in the same framework (the original implementation was for Java).

\subsection{Verifying Fixes of Different Qualities (RQ1) }

\begin{figure}[t!]
\centering\includegraphics[scale=0.34]{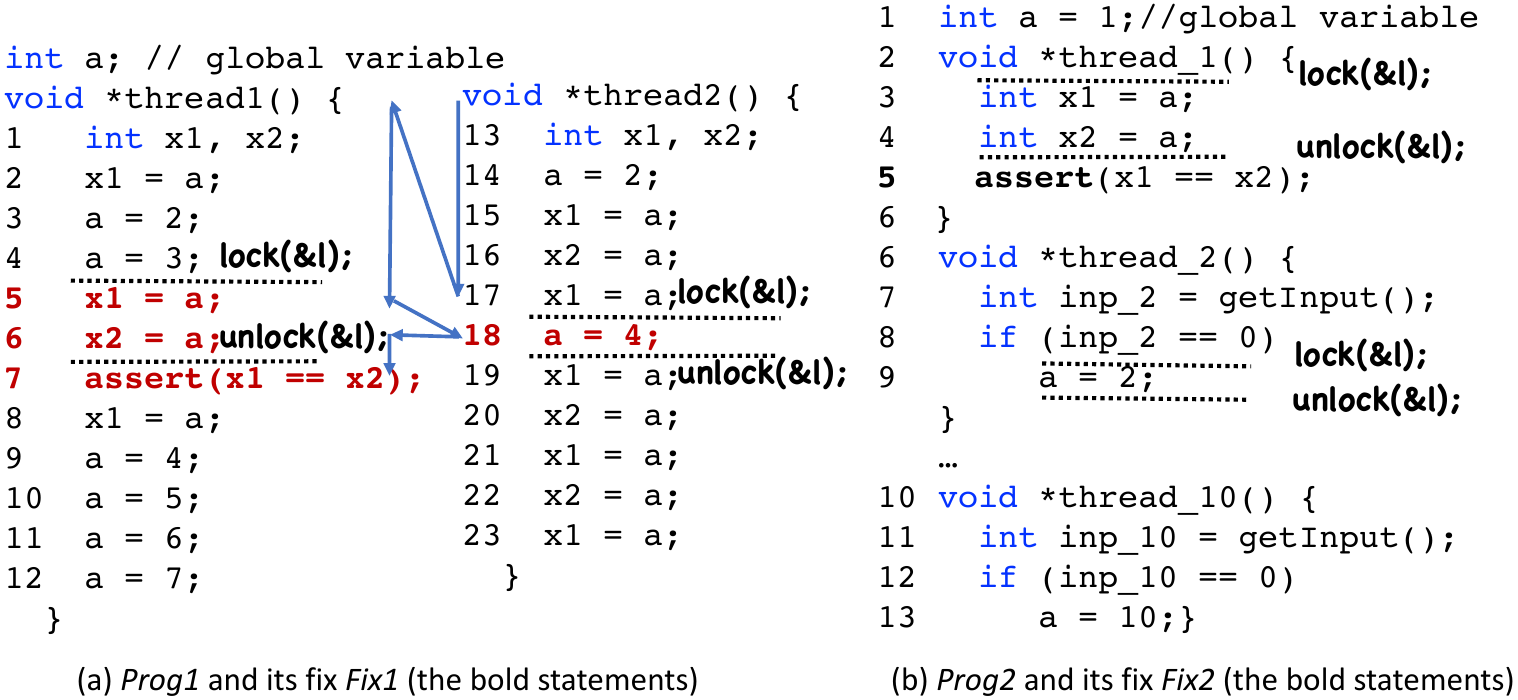}
\caption{\textit{$Fix_1$} and
\textit{$Fix_2$} belongs to \textit{InSuf}.}
\label{fig:fix1}
\end{figure}

According to the quality of the fixes, we classify the provided fixes into three types:

 \begin{itemize}
        \item  Suf: a correct fix, which can sufficiently and correctly enforce the intended atomicity.
	\item  InSuf: an incorrect fix, which cannot correctly enforce sufficient synchronization to the intended atomicity.
	\item  DL: an incorrect fix, which introduces new deadlocks.
 \end{itemize}
 \paragraph{\textbf{Prog1}} {\it Prog1} (Figure~\ref{fig:fix1}(a)) contains two threads, 
 executing  \texttt{thread1} and  \texttt{thread2} respectively, 
 and {\it Fix1} tries to fix the bug exposed by the buggy trace $\tau =\{13-17,1-5,18,6-7 \}$
 , as shown by the arrow lines. However, {\it Fix1} belongs to {\it InSuf}, 
 because there is still another assignment on variable $a$ at 
 line 14 in the buggy trace which may violate the intended atomicity after applying {\it Fix1}.

\paragraph{\textbf{Prog2}} {\it Prog2} (Figure~\ref{fig:fix1}(b)) contains ten threads $T_1-T_{10}$ 
with $T_i$ executing \texttt{thread\_i}. 
The buggy trace under input 
$inp\_2 = 0, inp\_3 = inp\_4 = ... = inp\_10 = 1$ 
only covers the assignment to \texttt{a} at line~9 of thread $T_2$. {\it Fix2} in Figure~\ref{fig:fix1} (b) belongs to 
{\it InSuf}, because it does not enforce the expected atomicity in 
\texttt{thread\_3}, . . . , and \texttt{thread\_10}.

\begin{figure}[t!]
\centering\includegraphics[scale=0.34]{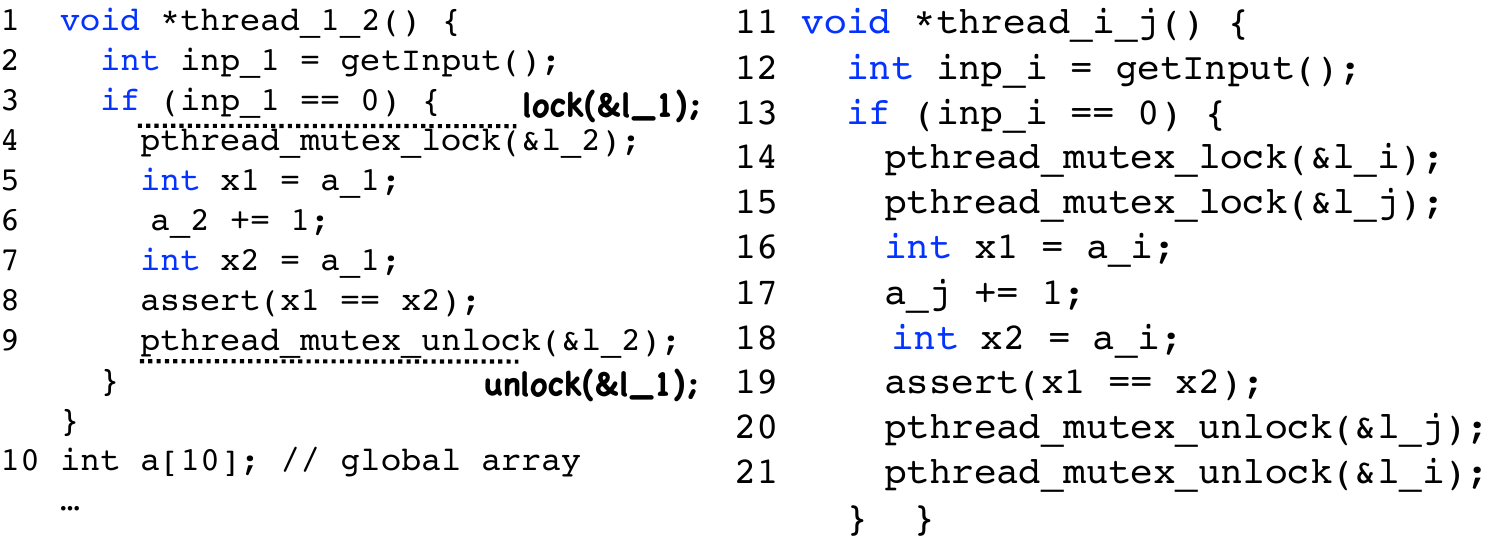}
\caption{\textit{$Prog_3$} and its fix \textit{$Fix_3$} (the bold statements),
which belongs to \textit{DL}.}
\label{fig:fix3}
\end{figure}

\paragraph{\textbf{Prog3}} {\it Prog3} (Figure~\ref{fig:fix3}) contains ten threads $T_1-T_{10}$. $T_i$
$(1 \leq i \leq 9)$ executes \texttt{thread\_i\_i+1} and $T_{10}$ executes \texttt{thread\_10\_1}. 
With input $inp\_1 = inp\_10 = 0, inp\_2 = inp\_3 = ... = inp\_9 = 1$, 
an atomicity violation happens because the expected atomicity 
unit between lines 5 - 7 is interrupted by line 17 of \texttt{thread\_10\_1}.
According to the programmer’s intention, 
the access on variable $a_i$ should be protected by lock $l_i$, 
thus {\it Fix3} adds a lock-unlock pair in 
\texttt{thread\_1\_2} to protect the access of $a_1$. 
Unfortunately, {\it Fix3} belongs to {\it DL} and introduces a deadlock, 
identified under input $inp\_1 = inp\_2 = inp\_3 = ... = inp\_9 = inp\_10 = 0$.

The steps \verifix\ uses to detect deadlocks are as follows: first, \verifix\ calculates the potential for deadlocks based on the explored traces using Algorithm~\ref{alg:potentialDL}, thereby identifying an unsafe cycle that may trigger a potential deadlock. However, Algorithm~\ref{alg:potentialDL} provides a coarse-grained estimation and may result in false positives. For example, if thread \texttt{thread\_1\_2} is the parent of other threads and has already released $l\_1$ and $l\_2$ before the other threads execute, a deadlock will never occur. Therefore, once Algorithm~\ref{alg:potentialDL} detects a suspicious deadlock, further constraints must be applied according to Algorithm~\ref{alg:checkdl} to determine whether there exists an input and scheduling combination that can trigger the deadlock. Ultimately, Algorithm~\ref{alg:checkdl} combines the deadlock constraints and trace constraints, then utilizes a constraint solver to determine whether there exists a program input and thread scheduling combination that can trigger the deadlock.
\paragraph{\textbf{Prog4}} Our motivating example in Figure~\ref{fig:movitation} 
and its three fixes, including two incorrect fixes \textit{fix1}, \textit{fix2}, 
renamed to \textit{Fix4}, \textit{Fix5} in Table~\ref{tab:myProg}, 
respectively. 
Specifically, \textit{Fix4} belonging to types {\it InSuf} and {\it DL}.

\begin{table*}
  \centering
  \caption{Results of comparing \verifix with \swan\ on Prog1 $\sim$ Prog4 }
  \label{tab:myProg}
  \begin{tabular}{rrrrrrrrrrr}
    \toprule
    \multirow{2}{*}{Program}&\multirow{2}{*}{Threads} & \multirow{2}{*}{Fix} & \multirow{2}{*}{Type} & \multicolumn{3}{c}{Swan} & \multicolumn{4}{c}{VeriFix} \\
  & & & & Run &Time(s)  & Success? & Run & Avg. Instr &Time(s) & Success? \\
    \midrule
    $Prog1$ &2 &Fix1 &InSuf   &2&0.21  &$\times$ &2   &9908 &0.07 &$\checkmark$\\
    $Prog2$ &10&Fix2 &InSuf   &0&0.03  &$\times$ &4   &11929&0.25 &$\checkmark$\\
    $Prog3$ &10&Fix3 &DL      &0&0.02  &$\times$ &280 &11901&18.5 &$\checkmark$\\
    $Prog4$ &3 &Fix4 &InSuf+DL&1&0.11  &$\checkmark/\times$ &1 & 11894&0.13&$\checkmark/\checkmark$\\
    $Prog4$ &3 &Fix5 &InSuf   &1&0.03  &$\times$ &1  &9820 &0.12  &$\checkmark$\\
    \bottomrule
  \end{tabular}
\end{table*}

\begin{table}
  \centering
  \caption{
Results of comparing \verifix\ with SPDOffline on deadlock repairs. The execution time of \verifix\ only includes the sum of the execution times of Algorithm~\ref{alg:potentialDL} and Algorithm~\ref{alg:checkdl}, given the same input traces as SPDOffline.}
  \label{tab:deadlock}
  \begin{tabular}{rrrrrr}
    \toprule
    \multirow{2}{*}{Program}&\multirow{2}{*}{Fix} & \multicolumn{2}{c}{SPDOffline} & \multicolumn{2}{c}{VeriFix} \\
  &  & Time(s)  & Success? &Time(s) & Success? \\
    \midrule
    $Prog3$ &Fix3 &0.1   &$\checkmark$ &0.28 &$\checkmark$\\
    $Prog4$ &Fix4 &0.04  &$\checkmark$ &0.15 &$\checkmark$\\
    Account &PFix &0.13  & ×           &0.21 &$\checkmark$ \\        
    \bottomrule
  \end{tabular}
\end{table}

 Table~\ref{tab:myProg} compares the results of verifying 
 Prog1 $\sim$ Prog4 by \verifix\ and \swan. We also present the average number of instructions executed by the path explorer during each run of \verifix.  
 Overall, \verifix\ successfully identified all incorrect fixes, as shown by $\checkmark$. 
 However, \swan\ made incorrect classification on {\it Fix1}$\sim$ {\it Fix5}. 
{\it Prog1} exposes one defect of \swan\ that it may miss 
the atomicity violations.  
For example, it ignored the fact that line 14 may violate the expected atomicity. 
For {\it Prog2} and {\it Prog3}, \swan\ even did not identify any suspicious atomicity violation. 
Thus, 
no extra execution was executed, 
as described by 0 under the column Run. 
For {\it Prog4}, \verifix\ identified two violations and one deadlock while verifying {\it Fix4} and 
{\it Fix5}. However, \swan\ only identified one violation.

Table~\ref{tab:deadlock} presents the results of \verifix\ and SPDOffline in detecting Fix3 and Fix4 in the first two columns. Since SPDOffline lacks the ability to explore traces, we provided it with the traces identified by \verifix\ that can trigger concurrency bugs. SPDOffline serves the same purpose as Algorithm~\ref{alg:potentialDL} and Algorithm~\ref{alg:checkdl}. The time shown for \verifix\ in the table only includes the time for Algorithm~\ref{alg:potentialDL} and Algorithm~\ref{alg:checkdl}. From the table, we can see that both \verifix\ and SPDOffline successfully identified the deadlocks.



\subsection{Evaluation on Real Bugs Caused by Insufficient Synchronization (RQ2)}

\begin{table*}
  \centering
  \caption{Results of comparing \verifix\ with \swan\ on a set of bugs caused by insufficient synchronization. }
  \label{tab:my-table}
  \begin{tabular}{rrrrrrrrrr}
    \toprule
    \multirow{2}{*}{CVE ID} & \multirow{2}{*}{Program} & \multirow{2}{*}{Type} & \multicolumn{3}{c}{Swan} & \multicolumn{4}{c}{VeriFix} \\
    & & & Run & Time(s) & Success? & Run &Avg. Instr & Time(s) & Success? \\
    \midrule
    CVE-2015-7550&Linux-4.3.4 &InSuf&1&0.47&\checkmark& 1 &11026&0.21 &\checkmark\\
    CVE-2016-1972&Firefox-45.0&InSuf&8&0.83&	×     &	4 &12122&0.41 &\checkmark\\
    CVE-2016-7911&Linux-4.6.6 &InSuf&1&0.50 &\checkmark&1 &12017&0.29 &\checkmark\\
    CVE-2016-9806&Linux-4.6.3 &InSuf&1&0.21&\checkmark&	1 &12040&1.12 &\checkmark\\
    \bottomrule
  \end{tabular}
\end{table*}

\begin{figure}[t!]
\centering\includegraphics[scale=0.75]{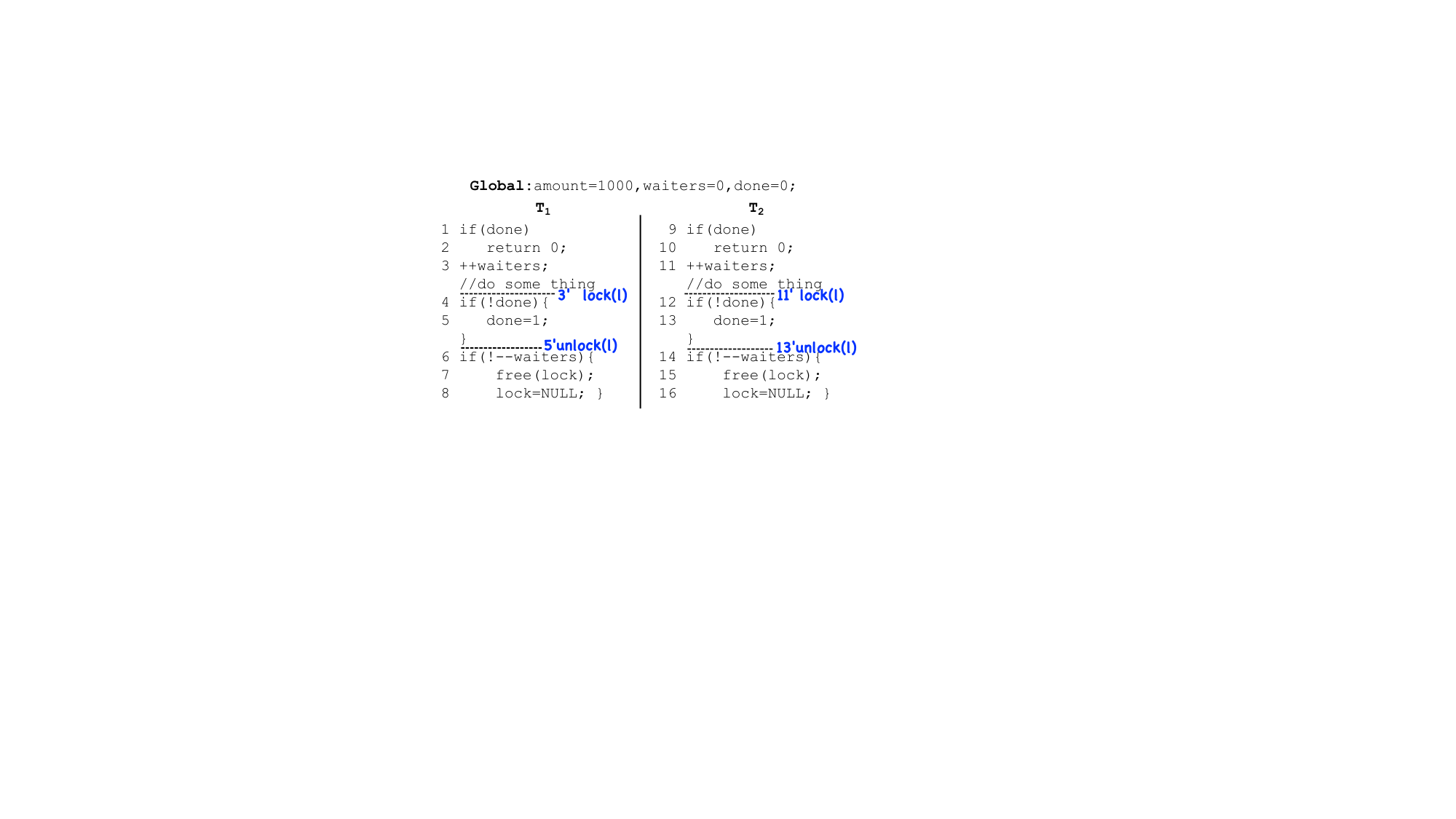}
\caption{The Double-Free bug triggered by trace $\tau_0=\{1,3,9,11,4,12,5,13$ $,6,14,15,7\}$ in \texttt{CVE-2016-1972}.}
\label{fig:cve}
\end{figure}

This subsection presents our evaluation results of \verifix\ on a set of real bugs 
from~\cite{cai19} caused by insufficient synchronization(shown in Table~\ref{tab:my-table}). 
Overall, \verifix\ successfully verifies all the insufficient synchronizations, 
while \swan\ failed to detect one of them. Furthermore, in most cases, 
\verifix\ terminated within less time and executions than \swan. Specifically, 
\verifix\ identified the insufficient synchronization in \texttt{CVE-2016-1972} after 
4 executions, but \swan\ failed after exploring all identified suspicious atomicity violations. 

For \swan, it failed when attempting to trigger the atomic violation in 
\texttt{CVE-2016-1972}, because it was unable to generate a feasible 
trace to trigger the atomic violation. 
Figure~\ref{fig:cve} presents the Double-Free bug triggered by trace 
$\tau_0=\{1, 3, 9,11,4,12,5,13,6,14,15,7\}$ in \texttt{CVE-2016-1972}. 
The root cause of this atomicity violation is $e_1$ and $e_5$ are interleaved by $e_9$. 
As shown in Figure~\ref{fig:cve}, the Bold blue statements constitute the fix for the triggered bug. 
\swan\ then identified 3 suspicious atomicity violations: $(e_6,e_{14},e_{15},e_7)$, 
$(e_9,e_5,e_{12})$ and $(e_1,e_{13},e_5)$. For $(e_6,e_{14},e_{15},e_7)$, \swan\ tried 
to generate trace $\tau_1=\{1,3,3',4,5,5',9,11,11',12,13,13',6,14,15,7\}$ to expose the violation. 
However, $\tau_1$ is infeasible since $e_9$ reads the value written by $e_5$, which is 1. 
Actually, \swan\ cannot ensure the reachability of each event, and 
it performed similarly on the other two suspicious atomicity violations. 
In contrast, \verifix\ uses path constraints $\Phi_{pc}$ and $\Phi_{rw}$ to 
guarantee the reachability of events. As a result, \verifix\ always generates 
feasible paths. 

\subsection{Evaluation on Fixes Generated by APR (RQ3)} 
Automated program repair (APR)~\cite{Goues19} is an emerging technology 
for automatic bug fix via search, semantic reasoning and learning. 
We selected a set of representative concurrent repair tools,
$\alpha$Fixer~\cite{Cai:17}, PFix~\cite{Lin18}, Hippodrome~\cite{Costea21}, 
to evaluate the efficiency of \verifix\ on verifying the fixes generated by APR tools.  
Since $\alpha$Fixer is not open source, and PFix and Hippodrome are for java. 
So we reproduced the tools for C/C++ based on their principles. 
Specifically, the input of $\alpha$Fixer includes 
the area of code that needs to be synchronized which 
may not be easy to obtain accurately in practice. 
We use \swan\ to identify suspicious violations.
The code region associated with the highest priority score corresponds to the suspicious violation location.
We then use the code area associated with the highest priority score corresponds to the suspicious violations as input to the $\alpha$Fixer.
All the used benchmarks are collected from the existing works on 
concurrent program testing~\cite{Yu:2009,Farchi:2003,Beyer19}.
\textit{Account} and \textit{Airline} are from the 
IBM ConTest benchmarks~\cite{Farchi:2003}. 
\textit{BoundedBuffer} is a shared bounded buffer implementation, 
\textit{Pfscan} is a parallel file scanner, and \textit{Aget} is a
multi-threaded download accelerator utility. \textit{ Apache} is an open-source HTTP server.
\textit{StringBuffer} is a C++ implementation with a bug in the Java JDK1.4 library, 
from prior work~\cite{Yu:2009}. 
The comparison results between \verifix\ and \swan\ are as shown in Table~\ref{tab:result-APR}.
Overall, \verifix\ identified all the insufficient or 
incorrect fixes within five runs. However, \swan\ failed to 
identify 4 incorrect atomicity violation fixes. 
Next, we will provide a more detailed description of several interesting benchmarks.

\paragraph{\textbf{Account}}

\begin{figure*}[t!]
  \centering
  \includegraphics[scale=0.7]{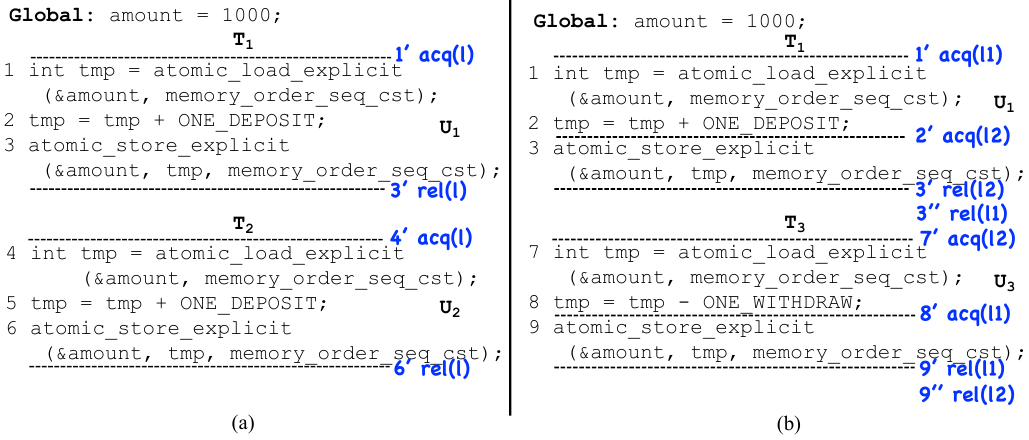}
  \caption{(a) An atomicity violation in Account. (b) PFix introduces a 
  deadlock while fixing an atomicity violation.}
  \label{fig:account}
\end{figure*}

Account is a program that simulates the storage and withdrawal of Bank Accounts.
As shown in Figure~\ref{fig:account}, 
it contains three threads ($T_1 \sim T_3$) 
and a shared variable $amount$.  
Thread $T_1$ and $T_2$ make deposits to the bank, and thread $T_3$ makes withdrawals.
However, the final amount will mismatch if work unit $U_1$ is interleaved by work unit $U_2$ following 
trace $\tau_0=\{1,4,2,3,5-9\}$, which manifests an atomicity violation. For $U_1$ and $U_2$, 
$\alpha$Fixer generates the fix represented in bold blue in Figure~\ref{fig:account}. 
For \verifix, it checks all possible atomicity violations along a trace, and thus 
it immediately identifies an atomicity violation introduced by work unit $U_3$ along $\tau_0$ 
after exploring the first execution. 
For \swan, it will not identify the generated fix is insufficient, since 
it will not identify any suspicious violations involving the work units (such as $U_3$) 
apart from $U_1$ and $U_2$. 


As reported in~\cite{Lin18}, PFix may introduce deadlock 
while fixing an atomicity violation. 
Specifically, when $e_1$ and $e_3$ was interleaved by $e_9$ following 
trace $\tau_0=\{1,2,7-9,3,4-6\}$, an atomicity violation was manifested. 
PFix generated a corresponding fix that inserted \textit{lock-unlock} pairs of lock $l_1$ around  
line~1 and line~3 as well as around line~9. With this fix, PFix then discovered 
that the interleaving $(e_7, e_3, e_9)$ also triggered the same atomicity violation. 
Thus, PFix introduced another lock $l_2$ to prevent this interleaving. 
Unfortunately, the two consecutive fixes (the bold black statements in Figure~\ref{fig:account}) caused a deadlock. 
\verifix\ successfully generated a schedule to trigger the deadlock after it 
identified the \textit{unsafe cycle}: $T_1$ holds $l_1$ while waiting for $l_2$, 
and $T_3$ holds $l_2$ while waiting for $l_1$. 
However, \swan\ missed the deadlock, since it terminated immediately when 
no suspicious violation was identified.
Additionally, as shown in Table~\ref{tab:deadlock}, SPDOffline failed to detect the deadlock. This is because SPDOffline imposes stricter constraints than \verifix. Specifically, to ensure the feasibility of events in the predicted trace, SPDOffline requires that all read events maintain the same read-write relationships as in the original trace, whereas in fact, some read events do not affect the reachability of the predicted trace. In contrast, \verifix\ only guarantees the feasibility of the path constraints in the predicted trace.


 \paragraph{\textbf{BoundedBuffer}}

 \begin{figure}[t!]
\centering\includegraphics[scale=0.6]{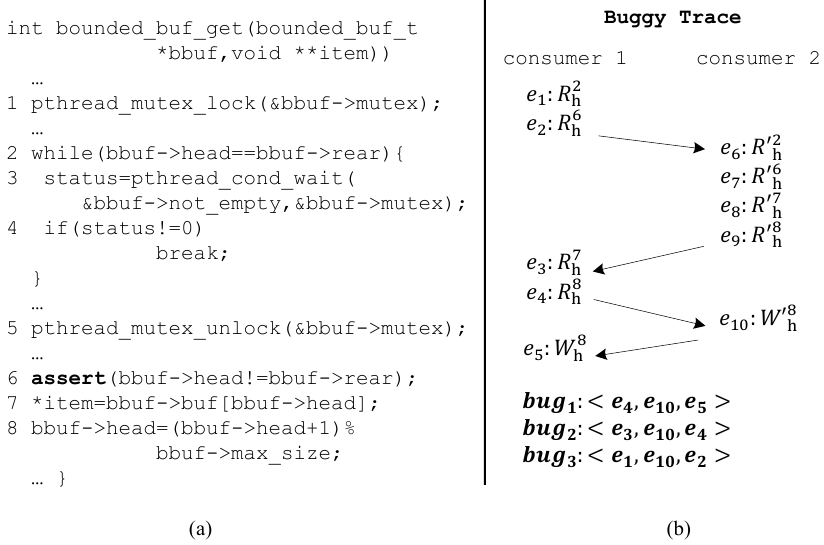}
\caption{(a) The simplified code snippet of the consumer. 
(b) The initial buggy trace, containing only the \texttt{read/write} events 
on variable $bbuf->head$, abbreviated as $h$.}
\label{fig:bbuf}
\end{figure}

BoundedBuffer is a C++ implementation of multi-threaded bounded buffer with semaphores. 
Our test driver contains 4 threads, including one parent thread, one producer, and two consumers.
The left part of Figure~\ref{fig:bbuf} shows the code snippet for the consumer which
consumes an \texttt{item} and updates \texttt{head} to record the index of the next available item.
The right part shows the buggy trace\footnote{Note that, the trace omits the events that are not related
to errors.} involving two consumers, which exposes an atomicity violation when
event $e_{10}$ is executed between $e_4$ and $e_5$, as described by $bug_1$.
$\alpha$Fixer introduced a lock-unlock pair on lock $bbuf$ around line 8 to exclude such interleaving. 
We constructed the property $\phi=(R_h^8=WPre_h^8)$ manually, where $WPre_h^8$ represents 
the value of $bbuf$-$\textgreater head$ just before $e_5$ is executed.
\swan\ identified $bug_2$ but not $bug_3$, because it can
catch the suspicious atomicity violation $\{e_3, e_{10}, e_4\}$ but not 
$\{e_1, e_{10}, e_2\}$. Specifically, it constrains 
$e_{10}$ be executed after $e_2$ as in the buggy trace. 

For \verifix, it first identified $bug_2$ by detecting that the value stored in 
\texttt{item} at line 7 is inconsistent with the expected one, when 
$e_{10}$ is executed between $e_3$ and $e_4$. After $bug_2$ is fixed by 
adding a lock-unlock pair around lines 7 and 8, we reran \verifix\ and then 
found $bug_3$, which triggers an assertion failure at line 6, when both consumers arrive 
at line 6 after checking that \texttt{bbuf} contains one element and 
$e_{10}$ is executed before $e_2$ (\texttt{consumer2} consumed the only element). 
In total, \verifix\ explored \texttt{5} executions to identify two atomicity violations, 
one in the first execution and the other in the fourth execution. 

\begin{table*}
  \centering
  \caption{Results of comparing \verifix\ with \swan\ on incorrect Fixes Generated by APR.}
  \label{tab:result-APR}
  \scalebox{0.85}{
  \begin{tabular}{rrrrrrrrrrr}
    \toprule
    \multirow{2}{*}{Program} & \multirow{2}{*}{APR} & \multirow{2}{*}{Threads} &\multirow{2}{*}{Type}&\multicolumn{3}{c}{Swan} & \multicolumn{4}{c}{VeriFix} \\
    & & & &Run & Time(s) & Success? & Run & Avg. Instr &Time(s) & Success? \\
    \midrule
    Account&$\alpha$Fixer      &3 &InSuf&0&0.03      &×&1         &12119     &0.45        &\checkmark \\
    Account&PFix             &3 &DL   &0&0.03        &×&1         &12210     &0.48        &\checkmark \\
    Pfscan& $\alpha$Fixer      &3 &InSuf&2&5.20      &\checkmark  &2         &21562       &8.20   &\checkmark\\
    BoundedBuffer&$\alpha$Fixer&4&InSuf&1/2&0.91/1.13&\checkmark/×&1/4       &13035/14562 &0.82/2.44 &\checkmark/\checkmark\\
    Aget         &PFix       &3&InSuf&3  &1.23     &×&5           &149364    &3.51      & \checkmark         \\
    StringBuffer &Hippodrome   &2&Insuf&1  &1.19     &\checkmark&1&12868      &1.03       &\checkmark \\
    Twostage     &Hippodrome   &2&InSuf&1  &0.26     &\checkmark&1&12058     &0.21       &\checkmark\\
    Reorder      &Hippodrome   &5&InSuf&1  &0.25     &\checkmark&1&12005     &0.22       &\checkmark\\
    Wronglock    &PFix       &4&InSuf&1  &1.02     &\checkmark  &2&12031     &0.40         &\checkmark\\
    Queue        &Hippodrome   &2&InSuf&1  &0.32     &\checkmark&1&11997     &0.21        &\checkmark\\
    Circular\_buffer&Hippodrome&2&InSuf&2  &0.18     &\checkmark&1&12067     &0.15        &\checkmark\\
    Bubblesort   &$\alpha$Fixer&2&InSuf&1  &0.21     &\checkmark&2&12112     &0.39        &\checkmark\\
    Stack        &Hippodrome   &2&InSuf&2  &0.32     &\checkmark&1&11976     &0.21        &\checkmark\\
    Apache-25520&$\alpha$Fixer &2&InSuf&1  &0.28     &\checkmark&1&13677     &0.22        &\checkmark\\
    \bottomrule
  \end{tabular}
  }
\end{table*}
 
 
 \paragraph{\textbf{Aget}}

 Aget is a multi-threaded wget-like 
 download accelerator utility. 
 A buggy trace exposes an atomicity 
 violation because 
 of two related variables 
 are not updated in an atomicity work unit.
 As shown in Figure~\ref{fig:aget}, the update on variable $h.wthread$ and $h.bwritten$ should be updated atomically. However, 
 the work unit $U_3$ is interrupted by the 
 work unit $U_1$, which changes wthread and bwritten used by $U_3$. 
 Fix generated by PFix is to add a 
 lock-unlock pair on a new lock around both $U_1$ and $U_3$. 
 The fix can successfully prevent the bug in the buggy trace. 
 For this fix, \swan\ stopped without identifying any bug. 
 However, \verifix\ found that this atomicity violation could 
 happen again but with a different input. Specifically, 
 the new bug-triggering input can drive the execution 
 to download with STP protocol, which invokes 
 unit $U_2$ similar to $U_1$, resulting in the same atomicity violation.

\subsection{Evaluation on Scalability (RQ4)}
\label{sec:scalability}

When a correct fix is provided, \verifix\ cannot determine whether untested paths still contain atomicity violations or deadlocks, leading it to explore the entire schedule and input space and thus the patch verification problem becomes equivalent to the program verification problem.
Thus, for further 
evaluating the scalability of \verifix, we compared it with Deagle, 
the SV-COMP 2022 ConcurrencySafety winner~\cite{SVCOMP2022}, which is 
the state-of-the-art concurrent verification tool implemented based on CBMC. 
It is worth noting that Deagle can not detect deadlocks, but \verifix\ 
detects possible deadlocks on every reachable path.


In this group of experiments, we selected the benchmarks from work ~\cite{Yu:2009}, 
which are widely used by APR techniques and thus are representative. 
Since Deagle can only detect C programs, we excluded the C++ programs. 
For each benchmark, the correct fix was provided. 
In Deagle, we used the default setting described in~\cite{Deagle2022} 
with a constant unwinding limit of 2. Since \verifix\ sets the loop unfolding 
depth to 5, we also presented the results under the scenario that Deagle's unwinding limit 
is set to 5. Moreover, in order to avoid unfairness, we symbolized all the variables 
affected by the functions that Deagle cannot handle. 
In this group of experiments, we set the memory limit to 8GB.



\begin{table*}
  \centering
  \caption{Results of comparing \verifix\ with Deagle on correct Fixes. 'OOM' represents 'Out of memory', and
  'TO' represents 'Time Out'.}
  \label{tab:result-Deagle}
 \scalebox{0.88}{
  \begin{tabular}{crrrrrrrr}
    \toprule
    \multirow{2}{*}{Program} &\multicolumn{4}{c}{Time(s)} & \multicolumn{3}{c}{Memory(MB)} \\
   & Deagle(2)&Deagle(5)&\verifix &\verifix-P & Deagle(2)&Deagle(5) & \verifix & \verifix-P \\
    \midrule
    Aget          &-    &-       &27.00   &14.23 &OOM    &OOM    &174&289\\
    Transmission  &-    &-       &1.12    &1.00  &OOM    &OOM    &34&44\\
    Httrack-3.43.9&10.5 &-       &6.60    &6.55  &1833   &OOM    &84&105\\
    Apache21285   &0.70 &Abort   &1.07    &0.78  &40     &-      &43 &174\\
    Apache45605   &1.05 &6.01    &32.00   &15.86 &46     &520    &46&198\\
    Apache21287   &TO   &TO      &1.01    &0.81  &1024   &3789   &46&155\\
    Apache25520   &TO   &TO      &7.90    &4.27  &6758   &93     &76&287\\
    Cherokee0.9.2 &Abort&Abort   &0.32    &0.36   &-     &-      &104&104\\
    \bottomrule
  \end{tabular}
                  }
\end{table*}

Table~\ref{tab:result-Deagle} describes the experiment results of comparing 
\verifix\ with Deagle. 
Specifically, \textit{Deagle(2)}/\textit{Deagle(5)} shows the results when Deagle's unwinding limit is set to 
2/5, while \textit{VeriFix-P} and \textit{VeriFix} respectively presents the results of 
parallel and non-parallel execution for \verifix. 
Furthermore, \textit{Memory} measures the peak memory usage during runtime, 
\textit{OOM} represents \textit{Out of Memory} when the memory limit is reached during runtime, 
\textit{TO} represents \textit{Time Out} when the runtime exceeds 20 minutes, 
and \textit{Abort} represents an abnormal termination encountered during execution.

As outlined in Table~\ref{tab:result-Deagle}, \verifix\ successfully completed its execution 
within the given time budget for all the benchmarks.
Nonetheless, Deagle succeeded in only three instances with an unwinding limit of 2,  
primarily due to \textit{OOM} or \textit{TO} issues. 
This success rate dropped to just one instance when the unwinding limit was increased to 5. 
Additionally, it is not difficult to find that Deagle usually requires 
significantly more memory than \verifix, as \verifix\ only analyzes the constraints along a single path at a time rather than the constraints of the entire program.
A more specific observation reveals that the primary limiting factor for VeriFix's memory overhead is the memory consumed by KLEE (specifically during the runtime of the path explorer), which is when the memory peak occurs.

Table~\ref{tab:result-Deagle} also provides a comparison between the sequential and parallel implementations of \verifix. The parallel implementation achieved an average speedup of 2.01× compared to its sequential counterpart. While the task partitioning and thread dispatching in our parallel implementation inevitably introduced some additional overhead, it is important to note that parallelization also incurs extra memory overhead. On average, \texttt{VeriFix-P} consumed $1.23×$ more memory than \verifix.

\subsection{Discussion}
\label{sec:discussion}

Our experimental data provides sufficient evidence for the main characteristics of \verifix. 
First, it is insensitive to the original buggy trace. 
Different input traces may exercise different program paths 
and thus contain different non-synchronized events, so 
this characteristic is vital for fixing verification techniques. 
Second, whenever a new bug is identified, \verifix\ provides a concrete \textit{schedule+input} 
combination to deterministically reproduce the bug, which can help with the following debugging process. 
Third, the path exploration process of \verifix\ can be easily parallelized 
when sufficient computing resources are provided.
Next, we will discuss the potential threats of this work.


\subsubsection{Path Explosion}
The biggest challenge faced by \verifix\ is the path explosion problem. While \verifix\ can quickly expose bugs when the fix is incorrect or insufficient, it needs to verify all paths in the program when the fix is correct, which can lead to path explosion in practice. This issue becomes more pronounced as the number of threads increases, as the total number of paths to explore is the product of the number of paths in each thread. If each thread has \( n_i \) paths, the total number of paths to explore would be \( \prod_{i=1}^{k} n_i \), where \( k \) is the number of threads.
For example, when the fix in the program is correct, setting the number of threads for bubblesort to 3 results in an execution time of 1.05s, 49.3s for 5 threads, and no termination even after 1 hour with 7 threads. Similarly, for Pfscan, execution time is 34s with 3 threads, but the program doesn’t stop within 1 hour at 5 threads. This exponential path explosion limits \verifix's verification capacity.

One possible solution is to integrate our approach with a change-aware technique. Instead of blindly exploring all unexplored paths, we could leverage static analysis to exclude paths that are irrelevant to the patch verification problem. We plan to incorporate these enhancements in future versions of \verifix.

Another potential solution is to reduce the number of paths to explore by leveraging the equivalence between different threads. For example, if the code bodies of threads \( T_1 \) and \( T_2 \) (which contain paths p1 and p2, respectively) correspond to identical code in the program under test, and both threads share the same initial setup, then these two threads are equivalent. In such cases, the program behaviors detected by testing \( p1 \)-\( \overline{p2} \), \( \overline{p2} \)-\( p1 \) would be identical. Of course, this approach is only applicable when equivalent threads exist.

\subsubsection{Modeling synchronization}
Another limitation is that our work, like \cite{Huang:MCR}, models synchronization constraints by considering only lock acquisition and release, as well as wait and notify, without taking into account other synchronization primitives. We will refine the modeling of different synchronization primitives in future work.

\subsubsection{Other Limitations}

We set the loop unfolding depth in KLEE to 5, and there may be verification failures caused by this setting in practical cases, although we have not encountered such issues so far. Additionally, incomplete modeling in KLEE could also lead to failure in triggering bugs. Furthermore, Z3 may fail to solve complex constraints, which is another potential risk that could compromise the soundness of our system.
Additionally, \verifix\ does not support dynamic code modifications, and the lack of support for just-in-time compilation scenarios may limit its broader applicability.

\subsubsection{External Validity} 
The evaluation of \textsf{VeriFix} was conducted on a set of real-world C/C++ programs. While these programs represent typical concurrent software, the results may not fully generalize to all concurrent systems. Future work will extend the evaluation to a broader range of software systems.

\section{Other Related Work}
\label{sec:related}

To our best knowledge, \swan~\cite{ShiHCX16} is the only 
technique that verifies synchronization fixes for atomicity violations.
However, many automated fixing techniques 
have been proposed to automatically generate patches to fix
real-world atomicity violation bugs~\cite{Liu:2012,Jin:2012,Liu:2014,LiuHao16,Jin:2011,Costea21,Cai:17,Lin18,Wang2020}. 
Unfortunately, it is usually impossible for them to obtain complete and accurate
information of a bug. So it is often a challenge for these 
techniques to generate the correct fix at the first attempt.

The key idea of PFix~\cite{Lin18} is to systematically fix concurrency bugs 
by inferring locking policies from failure-inducing memory-access patterns. 
However, aside from the scalability limitations of JPF~\cite{Willem04}, 
PFix may face challenges in accurately identifying the memory access patterns that lead to bugs.
Hippodrome~\cite{Costea21} repairs data races identified by RacerD~\cite{RacerD}, 
Facebook’s static concurrency analyzer for Java, by changing mutexes of Java synchronized blocks. 
However, the solution for fixing atomicity violations is formulated as a heuristic, it does not offer any guarantees. 
$\alpha$Fixer~\cite{Cai:17} firstly analyses the lock acquisitions of an 
atomicity violation. Then it either adjusts the existing lock scope or 
inserts a gate lock. However, it assumes that the areas of source code 
that need to be synchronized are known which may not be possible in practice. 
 


Some stateless model checkers (SMCs)~\cite{Huang:MCR,DC,RF-SMC,SMC-TSO} aim to mitigate the combinatorial explosion of interleavings in concurrent program verification, which grows exponentially. These tools are often combined with Dynamic Partial-Order Reduction (DPOR)~\cite{Flanagan:POR} to reduce the thread interleaving space that needs to be explored. DPOR records actual events during exploration and leverages this information on-the-fly. It guarantees the exploration of at least one trace in each Mazurkiewicz equivalence class~\cite{mazurkiewicz} when the state space is acyclic and finite. However, these approaches focus solely on exploring the thread interleaving space and do not consider the exploration of the program’s input space.

Some CBMC-based tools (e.g., Deagle~\cite{Deagle2022}) are not only effective in exploring the existing interleaving space but also in efficiently exploring the program's input space. CBMC~\cite{CBMC2004,CBMC2014} is a tool that detects assertion violations in C programs or proves the safety of assertions under a given bound. It works by translating the C program into a Conjunctive Normal Form (CNF) formula and then checking the satisfiability of this formula within a specified bound using an SMT solver like Z3. Apart from Deagle~\cite{Deagle2022}, there are several other well-known CBMC-based techniques for concurrent program verification. For example, Alglave et al.~\cite{alglave2013, alglave2010} use a partial order relation on memory events to represent possible executions caused by thread interleaving. All partial order constraints are encoded into a formula and queried through a dedicated solver for satisfiability.
These technologies are always memory intensive~\cite{kokologiannakis2017}.
Our technology differs from these in that we analyze only one path at a time, attempting to detect whether atomicity violations exist along that path, without requiring the entire program's state space simultaneously.

Regression testing for concurrent programs usually check the behavior
affected by code changes~\cite{Terragni:2015, Tucek:2009,
Jagannath:2011, Viennot:2013, Yin:2011}.
They are also known as incremental testing or delta testing. These methods mainly concentrate on
how to speed up the regression testing for concurrent programs.
~\cite{Jagannath:2011} proposes a change-aware technique that prioritizes the
exploration of schedules in a multithreaded regression test.
~\cite{Viennot:2013} proposes a mutual replay technique that can allow a recorded
execution of an application to be replayed with a patched version of the
application to validate if the patches successfully eliminate vulnerabilities.
The ideas of these techniques can be
integrated into \verifix\ to prune the error-free space or give higher
priority to those execution paths that are likely to expose bugs.

Researchers have proposed many approaches to encode traces as constraints, 
and then detect concurrency bugs by solving them~\cite{Huang:2014, Kahlon:2010,
Smaragdakis:2012,Gupta:2015,Huang:2015,Deshmukh:2011}.
CLAP~\cite{Huang:2013} proposes a concurrency bug reproduction approach 
through encoding the buggy trace as constraints and generating a concrete schedule 
to reproduce the encountered concurrency bug.  
Con2Colic~\cite{con2colic} uses concolic execution to generate schedules and 
inputs for testing concurrent programs. 
~\cite{samak2016} synthesizes programs 
to trigger violations of existing assertions or exceptions. 
However,
it lacks systematic path exploration capabilities. 
When exploring new paths, 
it only attempts to change the path 
constraints of a single thread, without considering the mutual impact of changing path constraints between threads.
Deshmukh~\cite{Deshmukh:2011} proposes 
an approach to detect deadlocks caused by
incorrect usage of libraries, by combining static analysis 
with a symbolic encoding scheme for tracking lock dependencies. 
Christian models deadlock prediction as a general decision problem. 
The proof of decidability maps the decision problem to an 
equivalent constraint problem that can be solved by an SMT-solver~\cite{deadlock}.
More recent work~\cite{deadlock-1} introduces a new class of predictable deadlocks, called synchronization-preserving deadlocks, for deadlock detection. However, these algorithms cannot generate execution traces, and their ability to detect deadlocks relies on input traces provided by other tools.

\section{Conclusion}
\label{sec:conc}


We have presented a new technique, \verifix, for verifying the 
correctness of atomicity violation fixes in concurrent programs.
To our best knowledge, it is the first technique that 
can systematically verify if a fix enforces sufficient 
synchronization in the whole schedule and input space without introducing any new deadlocks. 
We have implemented and evaluated \verifix\ 
on a collection of real-world C/C++ programs.
The result shows that \verifix\ significantly 
outperforms the state-of-the-art techniques.

\bibliographystyle{IEEEtran}
\bibliography{references}

\end{document}